\documentclass[11pt,draftcls,peerreview,onecolumn]{IEEEtran}

\usepackage[dvips]{graphics}
\usepackage{graphicx}
\usepackage{times}
\usepackage{amsmath}
\usepackage{amssymb}
\usepackage{fancybox}
\newtheorem{property}{\bf Property}
\newtheorem{algorithm}{\bf Algorithm}
\newtheorem{lemma}{\bf Lemma}
\newtheorem{theorem}{\bf Theorem}

\newtheorem{problem}{\bf Problem}

\newcommand{\bee}{\begin{eqnarray}}
\newcommand{\eee}{\end{eqnarray}}
\newcommand{\be}{\begin{equation}}
\newcommand{\ee}{\end{equation}}

\newcommand{\al}[1]{\begin{align} #1 \end{align}}

\newcommand{\equ}[1]{\begin{equation} #1 \end{equation}}

\newcommand{\mb}{\mathbf}

\newcommand{\bs}{\boldsymbol}
\newcommand{\wt}{\widetilde}
\newcommand{\nnb}{\nonumber}

\newcommand{\psmin}{P_0^{\min}}

\newcommand{\qa}{{\bf a}}

\newcommand{\qe}{{\bf e}}

\newcommand{\qh}{{\bf h}}

\newcommand{\qr}{{\bf r}}
\newcommand{\qs}{{\bf s}}

\newcommand{\qu}{{\bf u}}
\newcommand{\qv}{{\bf v}}
\newcommand{\qw}{{\bf w}}
\newcommand{\qx}{{\bf x}}

\newcommand{\qz}{{\bf z}}

\newcommand{\qA}{{\bf A}}

\newcommand{\qD}{{\bf D}}
\newcommand{\qE}{{\bf E}}

\newcommand{\qG}{{\bf G}}

\newcommand{\qI}{{\bf I}}

\newcommand{\qR}{{\bf R}}

\newcommand{\qU}{{\bf U}}

\newcommand{\qZ}{{\bf Z}}

\begin{document}
%
\title{Optimal Cooperative Relaying Schemes for Improving Wireless Physical Layer Security}

\author{\IEEEauthorblockN{Jiangyuan Li, Athina P. Petropulu, and Steven Weber}\\
\IEEEauthorblockA{Department of Electrical and Computer Engineering\\
Drexel University, Philadelphia, PA 19104}}

\maketitle

\begin{abstract}
\footnote{This work has been supported by NSF under Grant CNS-0905425.
Parts of this work were reported in \cite{LiConf1}, \cite{LiConf2}.
}
We consider a cooperative wireless network in the presence of one of more  eavesdroppers, and exploit node cooperation for achieving physical (PHY) layer based security. Two different cooperation schemes are considered. In the first scheme, cooperating nodes retransmit a weighted version of the source signal in a decode-and-forward (DF) fashion. In the second scheme, while the source is transmitting, cooperating nodes transmit weighted noise to confound the eavesdropper (cooperative jamming (CJ)). We investigate two objectives, i.e., maximization of achievable secrecy rate subject to a total power constraint, and minimization of total power transmit power under a secrecy rate constraint. For the first design objective with a single eavesdropper we obtain expressions for optimal weights under the DF protocol in closed form, and give an algorithm that converges to the optimal solution for the CJ scheme; while for multiple eavesdroppers we give an algorithm for the solution using the DF protocol that is guaranteed to converge to the optimal solution for two eavesdroppers.  For the second design objective, existing works introduced additional constraints in order to reduce the degree of difficulty, thus resulting in suboptimal solutions. In this work, either a closed form solution is obtained, or algorithms to search for the solution are proposed. Numerical results are presented to illustrate the proposed schemes and demonstrate the advantages of cooperation as compared to direct transmission. 

\end{abstract}


\newpage

\section{\label{sec:1}Introduction}

Privacy and security issues play an important role in wireless networks.  Although security is typically addressed via cryptographic approaches, there have been several attempts at addressing security at the physical layer, following the pioneering work of \cite{Wyner}. Recently wireless physical (PHY) layer based security from a information-theoretic point of view has received considerable attention, e.g., \cite{Liang}-\cite{Liang:book}. The wiretap channel, first introduced and studied by Wyner \cite{Wyner}, is the most basic physical layer model that captures the problem of communication security. Wyner showed that when an eavesdropper's channel is a degraded version of the main channel, the source and destination can achieve a positive perfect information rate ({\em secrecy rate}). The maximal secrecy rate from the source to the destination is defined as the {\em secrecy capacity} and for the degraded wiretap channel is given as the difference between the rate at the legitimate receiver and the rate at the eavesdropper. The Gaussian wiretap channel, in which the outputs at the legitimate receiver and at the eavesdropper are corrupted by additive white Gaussian noise (AWGN), was studied in \cite{Hellman}. Along the same lines, the Gaussian MIMO wiretap channel was investigated and the secrecy capacity of the MIMO wiretap channel was established in terms of an optimization problem over all possible input covariance matrices \cite{Khisti}, \cite{Hassibi}. There have also been some recent works focusing on secrecy rates based on partial CSI or channel statistics \cite{Liang, Gopala, Bhargava}. In \cite{Bhargava}, the authors derived the ergodic secrecy capacity of Gaussian MIMO wiretap channel and showed that a circularly symmetric Gaussian input is optimal.

The secrecy rate is affected by   channel conditions between the source and the destination and also channel conditions between the source and the eavesdroppers. A low cost approach to increase the achievable secrecy rate by exploiting/mitigating channel effects is node cooperation via relays \cite{Tekin}-\cite{Aggarwal}. A two-stage cooperative approach was recently proposed in \cite{Dong1,Dong2}, and their extended version \cite{Dong3}. In \cite{Dong3}, the source first transmits locally to a set of trusted relays, and subsequently, the relays retransmit a weighted version of the signal that they heard (amplify-and-forward (AF)), or a weighted version of the decoded signal (decode-and-forward (DF)). Alternatively, the relays can transmit weighted noise to confound the eavesdropper while the source is transmitting (cooperative jamming (CJ)). In all cases, the objective is to select the weights so as to maximize the secrecy rate under total power constraints, or to minimize the total power under a secrecy rate constraint. The results in \cite{Dong1}-\cite{Dong3} contain sub-optimal weights for both a single eavesdropper and multiple eavesdroppers, due to the difficulty of solving the associated optimization problems. In particular, several criteria for sub-optimal weight design were proposed such as completely nulling out the message signal at all eavesdroppers for DF and AF, and completely nulling out the jamming signal at the destination for CJ. These sub-optimal weights may yield a reduction in the achievable secrecy rate or minimization of total power.

In this paper, we consider the same scenario and problem as in \cite{Dong1}-\cite{Dong3}, but focus on obtaining the optimal solution for the DF and CJ schemes. Obtaining the solution  for the AF scheme is a more difficult problem and will be addressed in future work. Exploiting certain properties of the objective functions and the constraints, enables us to either obtain closed form solutions, or, if a closed form solution is not possible, propose  algorithms to search for the solution. 

The remainder of this paper is organized as follows. The mathematical model is introduced in \S\ref{sec:2}. In \S\ref{sec:3} we derive the optimal relay weights  that maximize the achievable secrecy rate subject to a total power constraint in the presence of a single eavesdropper for the DF and CJ protocols, and multiple eavesdroppers for DF protocol. In \S\ref{PminSecRateCons} we study the optimal weights that minimize the total power under a secrecy rate constraint for the DF and CJ protocols. Numerical results in \S\ref{sec:5} are presented to illustrate the proposed solutions. Finally, \S\ref{sec:6} provides some concluding remarks.

\subsection{Discussion of related work}

Our work falls under the general scenario where the source-destination communication is aided by a relay or a helper. Relevant results include the work of \cite{Tekin},  where multiple users communicate with a common receiver  in the presence of an eavesdropper, and the transmit power allocation policy is determined that maximizes the secrecy sum-rate. In \cite{Lai}, a source, destination, eavesdropper and relay model is considered, in which the relay transmits a noise signal in order to jam the eavesdropper. The rate-equivocation region is derived to show gains and applicable scenarios for cooperation, with the equivocation denoting the uncertainty of the eavesdropper about the source message.

A generalization of \cite{Tekin} and \cite{Lai} was proposed in \cite{Tang}, in which the helper transmits signals from another source encoder. In \cite{Yuksel}, inner and outer bounds on the rate-equivocation region were derived for the four-node model for both discrete memoryless and Gaussian channels. In \cite{Aggarwal}, the secrecy rate of orthogonal relay and eavesdropper channels was studied.

Our work in this paper is different from the aforementioned works in the sense that we address the more general case of multiple relays and multiple eavesdroppers. Also, existing works primarily focus on rate-achieving relaying strategies. In our work, we consider pre-defined cooperative schemes without claiming that those schemes are optimal, and determine relay weight  and power allocation design that optimize the achievable secrecy rate subject to a power constraint, or minimize the transmit power subject to a secrecy rate constraint.

\subsection{Notation}

Upper case and lower case bold symbols denote matrices and vectors, respectively. Superscripts $\ast$, $T$ and $\dagger$ denote respectively conjugate, transposition and conjugate transposition. $\mathrm{Tr}({\mb A})$ denotes the trace of matrix $\mb A$. ${\mb A}\succeq 0$ means that ${\mb A}$ is a Hermitian positive semi-definite matrix. $\mathrm{rank}(\qA)$ denotes the rank of matrix $\qA$. $\|\qa\|$ denotes Euclidean norm of vector $\qa$. $\qI_n$ denotes the identity matrix of order $n$ (the subscript is dropped when the dimension is obvious). $f'(x)$ and $f''(x)$ denote  first- and second- order derivatives of $f(x)$, respectively.

\begin{figure}[hbtp]
\centering
\includegraphics{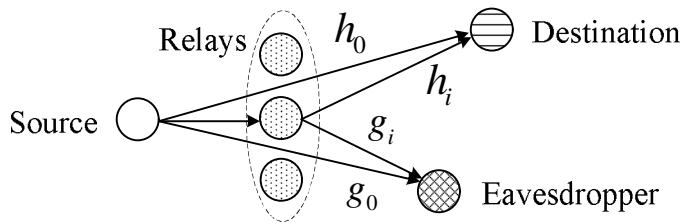}
\caption{System model: source $S$ wishes to communicate to destination $D$ in the presence of $J$ eavesdroppers, $E_1,\ldots,E_J$. The $N$ relays, $R_1,\ldots,R_N$ implement decode and forward (DF) or cooperative jamming (CJ), based protocols.  In each case the objective is to select the relay weights ($\qw$) and the source power ($P_s$) to maximize the achievable  secrecy rate subject to a total power constraint ($P_0$), or to minimize the total power constraint under a secrecy rate constraint.}
\label{fig:1}
\end{figure}

\section{\label{sec:2}System Model and Problem Statement}

We consider a wireless network model depicted in Fig.~\ref{fig:1}, consisting of one source node $S$, a set of $N$ relay nodes $(R_i,~i=1,\ldots,N)$, a destination node $D$, and a set of $J$ passive eavesdroppers $(E_j,~j=1,\ldots,J)$. The symbols used in the paper are listed in Table~\ref{tab:1}.

\begin{table}[hbtp]
\centering
\caption{Mathematical notation}
\label{tab:1}
\begin{tabular}{lll}
\hline
  $N$ & number of relays\\
  $J$ & number of eavesdroppers\\
  $P_0$ & total power (source power plus the relays' power)\\
  $P_s$ & transmit power at the source\\
  $\sigma^2$ & noise variance\\
  $h_0$ & baseband complex channel
gain  between the source and the destination\\
  $h_i$ & baseband complex  channel gain between the $i$th relay and the destination \\
  $a_i$ & baseband complex  channel gain between the source and the $i$th relay \\
  $g_{0j}$ & baseband complex channel gain between the source and the $j$th eavesdropper\\
  $g_{ij}$ & baseband complex channel gain between the $i$th relay and the $j$th eavesdropper\\
  $\qw$ & weight vector at the relays, $[w_1,\cdots,w_N]^T$\\
  $\qh$ & $[h_1,\cdots,h_N]^T$\\
  $\mb{g}_j$ & $[g_{1j},\cdots,g_{Nj}]^T$\\
  $\qR_h$ & $\qh\qh^\dagger$\\
  $\qR_g^j$ & $\mb{g}_j\mb{g}_j^\dagger$\\
  \hline\\
  {\it Notes}:& The index $j$ is dropped when $J=1$.\\
\end{tabular}
\end{table}

The source message  is uniformly distributed over the message set $\mathcal{W}= \{1,2,..., 2^{nR}\}$, which is transmitted in $n$ channel uses. Here, $R$ denotes the source rate (unit: bits per channel use) and the message has entropy $nR$ bits. A stochastic encoder at the source maps each message to a codeword from an alphabet of length-$n$. For the purpose of evaluating the achievable secrecy rate, we assume that the codewords used at the source are Gaussian. We consider a time division multiple access system, in which there are $n$ time units in each transmission slot. In a time unit, the average power of an encoded source symbol is normalized to unity.  The noise at any node is assumed to be zero-mean white complex Gaussian with variance $\sigma^2$. Each node is equipped with a single omni-directional antenna and operates in a half-duplex mode.

All channels are assumed to be flat fading. We assume that global channel state information (CSI) is available, including the eavesdroppers' channels. This corresponds to the cases where the eavesdroppers are active in the network and their transmissions can be monitored \cite{Bloch}. We should note  that there have been some recent works focusing on secrecy rates based on partial CSI or channel statistics (e.g., chapter 5 in \cite{Liang:book}, and \cite{Gopala}). Adapting the proposed work to cooperative schemes that uses partial CSI or channel statistics will be considered  in future work.

Similarly as in \cite{Yuksel:ITW07}, \cite{Dong3}, we assume that the source encoding scheme, the decoding methods at  destination and  eavesdroppers, and the cooperative protocol, are all public information.


Let us fix the relay weight vector $\qw$ and the source transmission power $P_s$. Then the following expressions give the rates with the destination and the eavesdroppers as a function of $\qw$, $P_s$, the noise $\sigma^2$, and the various channel gains. For the {\em DF-based protocol}, the  rate at the destination and the $j$th eavesdropper are, respectively
\al{
R_d&=\frac{1}{2}\log\left(1+\frac{P_s|h_0|^2+\qw^\dagger\qR_h\qw}{\sigma^2}\right),\label{CdDF}\\
R_e^j&=\frac{1}{2}\log\left(1+\frac{P_s|g_{0j}|^2+\qw^\dagger\qR_g^j\qw}{\sigma^2}\right)\label{CdDF2}
}
where the scalar factor $1/2$ is due to the fact that two time units are required in two stages. Here we have assumed an additional constraint, i.e., $P_s\ge \psmin$ where $\psmin$ is the minimum source power requirement for cooperative nodes to correctly decode the source message with high probability. We assume $\psmin$ is known a priori. For the {\em CJ-based protocol}, the rate at the destination and the $j$th eavesdropper are, respectively
\al{
R_d&=\log\left(1+\frac{P_s|h_0|^2}{\qw^\dagger\qR_h\qw+\sigma^2}\right),\label{CdJF}\\
R_e^j&=\log\left(1+\frac{P_s|g_{0j}|^2}{\qw^\dagger\qR_g^j\qw+\sigma^2}\right).\label{CdJF2}
}
For both the DF and CJ  protocols, the achievable secrecy rate in the presence of $J$ eavesdroppers is given by \cite{Liang:Allerton07}
\equ{
R_s=\max\{0, R_d-\max_{1\le j\le J}R_e^j\}.\label{Cs0}
}
In particular, when $J=1$, i.e., a single eavesdropper, the secrecy rate in (\ref{Cs0}) becomes \cite{Hassibi, Oggier}
\equ{
R_s=\max\{0, R_d-R_e\}.
}
We consider the practical case in which the system can be designed so that the secrecy rate is positive. In that case, the achievable secrecy rate can be rewritten as
\equ{
R_s=R_d-\max_{1\le j\le J}R_e^j.\label{Cs}
}

Achievability of the above rates can be shown based on
existing results for MIMO wire-tap
channels, such as    \cite{Khisti2}, \cite{Oggier},
\cite{TieLiu}, for one eavesdropper,
and \cite{Liang:Allerton07} for multiple eavesdroppers.

The CJ scheme can be viewed as a  $1 \times 2$ SIMO system, so that MIMO results are directly applicable. As discussed in \cite{Dong3}, for the DF scheme, MIMO results are applicable if we assume that  the received signal at destination/eavesdropper at time $i$ depends only on the relays' transmitted encoded signals at time $i$ (though a relay's transmitted signal at time $i$ depends on its received signal before time $i$). This is usually referred to as the ``memoryless relay channel''  \cite{Lai}, \cite{Yuksel}. For convenience, as in \cite{Dong3} we focus on case in which the relays use the same codewords as the source to re-encode the signal before transmission; in that case the rates of source-destination and source-eavesdropper links admit simple closed-form expressions \cite{Dong3}.


The problems addressed in this paper are described as follows.
\begin{problem}[maximize secrecy rate under power constraint using DF]
\label{P1}
Given total power (source plus relays) $P_0$, select source power $P_s$ and cooperative nodes' weights
$\qw$ to maximize the secrecy rate
\begin{equation}\label{obj1}
    \max_{P_s, \qw} \ R_s\quad \mathrm{s.t.} \ P_s+\|\qw\|^2=P_0, P_s\ge \psmin.
\end{equation}
where $R_s$ is given by (\ref{CdDF}), (\ref{CdDF2}) and (\ref{Cs}).
\end{problem}
The solution of Problem 1 for one eavesdropper is provided in Section \ref{ssec:3a}, and for multiple eavesdroppers in Section \ref{sec:4}.

\begin{problem}[maximize secrecy rate under power constraint using CJ]\label{P2}
Given total power (source plus relays) $P_0$, select source power $P_s$ and cooperative nodes' weights
$\qw$ to maximize the secrecy rate
\begin{equation}\label{obj1}
    \max_{P_s, \qw} \ R_s\quad \mathrm{s.t.} \ P_s+\|\qw\|^2=P_0.
\end{equation}
where $R_s$ is given by (\ref{CdJF}), (\ref{CdJF2}) and (\ref{Cs}).
\end{problem}
The solution of Problem 2 for one eavesdropper is provided in Section \ref{ssec:3b}.

\begin{problem}[minimize transmit power under secrecy rate constraint using DF]\label{P3}
Given the secrecy rate constraint $R_s^0$, select the source
power $P_s$ and cooperative nodes' weights, $\qw$, to minimize the
total power (source plus relays)
\al{
\min_{P_s, \qw}\
\left[P_0=P_s+\|\qw\|^2\right]\quad\mathrm{s.t.}\ R_s=R_s^0, P_s\ge \psmin \label{powermin}
}
where $R_s$ is given by (\ref{CdDF}), (\ref{CdDF2}) and (\ref{Cs}).
\end{problem}
The solution of Problem 3  is provided in Section \ref{Pminsec:3sub:1}.

\begin{problem}[minimize transmit power under secrecy rate constraint using CJ]\label{P4}
Given the secrecy rate constraint $R_s^0$, select the source
power $P_s$ and cooperative nodes' weights, $\qw$, to minimize the
total power (source plus relays)
\al{
\min_{P_s, \qw}\
\left[P_0=P_s+\|\qw\|^2\right]\quad\mathrm{s.t.}\ R_s=R_s^0 \label{powermin}
}
where $R_s$ is given by (\ref{CdJF}), (\ref{CdJF2}) and (\ref{Cs}).
\end{problem}
The solution of Problem 4  is provided in Section \ref{Pminsec:3sub:2}.

Before proceeding, we provide  Lemma \ref{lem:1} and \ref{lem:2},
which will be the basis of the results to follow. Please see
Appendix \ref{proofLem1} and \ref{proofLem2} for details.
\begin{lemma}\label{lem:1}
{\em Let $\qr$ and $\qs$ be (known) linearly uncorrelated vectors.
Let $\theta$ be the argument of $\qr^\dagger\qs$. Denote $\mathrm{i}=\sqrt{-1}$.
The matrix $\qr\qr^\dagger-\qs\qs^\dagger$ has only two nonzero
eigenvalues, i.e.,  $\eta_1>0$ and $\eta_2<0$, given by \equ{
\eta_1=\|\qr\|^2-|c_2||\qr^\dagger\qs|,\
\eta_2=\|\qr\|^2-|c_4||\qr^\dagger\qs|. } The corresponding
eigenvectors are \equ{
\qe_1=c_1(\qr+|c_2|e^{\mathrm{i}(\pi-\theta)}\qs),\
\qe_2=c_3(\qr+|c_4|e^{\mathrm{i}(\pi-\theta)}\qs) } where
$c_1=1/{\sqrt{\|\qr\|^2+|c_2|^2\|\qs\|^2-2|c_2||\qr^\dagger\qs|}}\,$,
$c_3=1/{\sqrt{\|\qr\|^2+|c_4|^2\|\qs\|^2-2|c_4||\qr^\dagger\qs|}}\,$,
$2|c_2| |\qr^\dagger\qs|
=\|\qr\|^2+\|\qs\|^2-\sqrt{(\|\qr\|^2+\|\qs\|^2)^2-4|\qr^\dagger\qs|^2}\,$,
and $2|c_4|
|\qr^\dagger\qs|=\|\qr\|^2+\|\qs\|^2+\sqrt{(\|\qr\|^2+\|\qs\|^2)^2-4|\qr^\dagger\qs|^2}\,$.}
\end{lemma}

\begin{lemma}\label{lem:2}
{\em Let $\mb{d}_1$ and $\mb{d}_2$ be (known) unit-norm vectors.  Let
$\phi\in (-\pi, \pi]$ be the argument of $\mb{d}_2^\dagger\mb{d}_1$,
$r=|\mb{d}_1^\dagger\mb{d}_2|$, and consider $0\le q\le 1$. The solution
of \al{ \max_{\qz} \
\qz^\dagger\mb{d}_2\mb{d}_2^\dagger\qz\quad\quad \mathrm{s.t.} \ \
\qz^\dagger\mb{d}_1\mb{d}_1^\dagger\qz=q, \ \|\qz\|=1\label{lem1op1}
}
is given by $\qz^\circ=c_1\mb{d}_1+c_2\mb{d}_2$, where
$c_2=\sqrt{(1-q)/(1-r^2)}$ and
$c_1= (rc_2-\sqrt{q}) e^{\mathrm{i}(\pi-\phi)}$. The maximum is
${\qz^\circ}^\dagger\mb{d}_2\mb{d}_2^\dagger\qz^\circ=1-(r\sqrt{1-q}-\sqrt{(1-r^2)q})^2$.}
\end{lemma}

\section{\label{sec:3}Secrecy Rate Maximization under Power Constraint}

In this section, we address Problems \ref{P1} and \ref{P2}.

\subsection{\label{ssec:3a}One Eavesdropper ($J=1$)}

In this subsection we address Problem \ref{P1} for the case of a single eavesdropper. Since $J=1$ we drop the superscript $j$ from $R_e^j$.

\subsubsection{\label{ssec:3a}DF-based protocol}

Problem \ref{P1} can be recast as
\al{
&\max_{P_s, \qw} \
\frac{1}{2}\log\left(\frac{\sigma^2+P_s|h_0|^2+\qw^\dagger\qR_h\qw}{\sigma^2+P_s|g_0|^2+\qw^\dagger\qR_g\qw}\right)
\label{MaxSecCapDFoneEaveOP}\\
&\mathrm{s.t.} \quad P_s\ge \psmin, \qw^\dagger\qw=P_0-P_s. \nnb
}

Denote $\qu_1=\mb{g}/\|\mb{g}\|$, $\qu_2=\qh/\|\qh\|$ and $\zeta=|\qu_1^\dagger\qu_2|$. Let $\theta \in (-\pi, \pi]$ be the argument of $\qu_2^\dagger\qu_1$. The solution of Problem \ref{P1} is given in the following theorem. The proof is given in Appendix \ref{proofTheo-solutionP1}.
\begin{theorem}
\label{solutionP1}
{\em The solution of (\ref{MaxSecCapDFoneEaveOP}) is given by
\equ{
P_s^\circ = \left\{\begin{array}{cl}
                     \psmin & \mathrm{if\ } J(\psmin)>J(P_0); \\
                     P_0 & \mathrm{else;}
                   \end{array}
\right.
}
\equ{
\qw^\circ = \left\{\begin{array}{cl}

c_1\qu_1+c_2\qu_2 & \mathrm{if\ } J(\psmin)>J(P_0); \\
                     0 & \mathrm{else}
                   \end{array}
\right.
}
where the function $J(P_s)$ is defined by
\equ{
J(P_s)=\frac{\sigma^2+P_s|h_0|^2+(P_0-P_s)\|\qh\|^2L(z(P_s))}{\sigma^2+P_s|g_0|^2+(P_0-P_s)
\|\mb{g}\|^2z(P_s)}\label{JPsDFone}
}
and
\al{
&L(z)\triangleq 1-(\zeta\sqrt{1-z}-\sqrt{(1-\zeta^2)z})^2\\
&z(P_s)=({B-\sqrt{B^2-4C}})/({2bC})-{a}/{b}\label{zPs}\\
&a=\sigma^2+P_s|g_0|^2,\nnb\\
&b=(P_0-P_s)\|\mb{g}\|^2,\nnb\\
&c=\sigma^2+P_s|h_0|^2+(P_0-P_s)\|\qh\|^2(1-\zeta^2),\nnb\\
&d=(P_0-P_s)\|\qh\|^2(1-2\zeta^2),\nnb\\
&f=2(P_0-P_s)\|\qh\|^2\zeta\sqrt{1-\zeta^2},\nnb\\
&B=({2a+b})/({a^2+a b}),\nnb\\
&C=\frac{f^2(2a+b)^2+4(ad+bc)^2}{4f^2(a^2+ab)^2+4(ad+bc)^2(a^2+ab)}\nnb\\
&c_2=\sqrt{P_0-P_s^\circ}\,\sqrt{(1-z(P_s^\circ))/(1-\zeta^2)}\nnb\\
&c_1=\sqrt{P_0-P_s^\circ}\,(\zeta c_2-\sqrt{z(P_s^\circ)})e^{\mathrm{i}(\pi-\theta)}.\nnb
}
}
\end{theorem}
{\em Remarks:} In particular, Theorem \ref{solutionP1} states that, depending on the relative values of $J(\psmin)$ and $J(P_0)$,
the optimal solution is either i) the source uses minimum power and the relay weights are a linear combination of the normalized relay and eavesdropper channel vectors, or ii) the source uses all the power and the relays are unused.

\subsubsection{\label{ssec:3b}CJ-based protocol}

Problem \ref{P2} is recast as
\al{
&\max_{P_s,\qw}\
\log\left(1+\frac{P_s|h_0|^2}{\qw^\dagger\qR_h\qw+\sigma^2}\right)
-\log\left(1+\frac{P_s|g_0|^2}{\qw^\dagger\qR_g\qw+\sigma^2}\right)\label{MaxSecCapCJoneOP1}\\
&\mathrm{s.t.}\ \ \qw^\dagger\qw=P_0-P_s, P_s\in [0,P_0].\nnb
}
By denoting $\qw=\sqrt{P_0-P_s}\,\qx$, $\qv_1=\qh/\|\qh\|$,
$\qv_2=\mb{g}/\|\mb{g}\|$, the problem of (\ref{MaxSecCapCJoneOP1}) can be rewritten as
\al{
&\max_{P_s,\qx}\
\log\left(1+\frac{P_s|h_0|^2}{(P_0-P_s)\|\qh\|^2\qx^\dagger\qv_1\qv_1^\dagger\qx+\sigma^2}\right)
-\log\left(1+\frac{P_s|g_0|^2}{(P_0-P_s)\|\mb{g}\|^2\qx^\dagger\qv_2\qv_2^\dagger\qx+\sigma^2}\right)\label{MaxSecCapCJoneOP1-b}\\
&\mathrm{s.t.}\ \ \qx^\dagger\qx=1, P_s\in [0,P_0].\nnb
}
Let $\qx$ be a feasible point. Denote $\qx^\dagger\qv_1\qv_1^\dagger\qx=z$, $z\in [0,1]$.
For fixed $z$, a larger $\qx^\dagger\qv_2\qv_2^\dagger\qx$ results in a larger
objective value. With this and from Lemma \ref{lem:2}, we know that the optimal
$\qx^\dagger\qv_2\qv_2^\dagger\qx$ equals $G(z)$ where
$G(z)\triangleq1-(\eta\sqrt{1-z}-\sqrt{(1-\eta^2)z})^2$,
$\eta=|\qv_1^\dagger\qv_2|$.
With these, we can rewrite the optimization of
(\ref{MaxSecCapCJoneOP1-b}) as
\al{
&\max_{P_s, z}\ \log\left(1+\frac{P_s}{(P_0-P_s)\alpha_1 z+\alpha_2}\right) -\log\left(1+\frac{P_s}{(P_0-P_s)\alpha_3G(z)+\alpha_4}\right)\label{MaxSecCapCJoneOP2}\\
&\mathrm{s.t.}\quad z\in [0,1], P_s\in [0,P_0]\nnb
}
where
$\alpha_1=\|\qh\|^2/|h_0|^2$, $\alpha_2=\sigma^2/|h_0|^2$,
$\alpha_3=\|\mb{g}\|^2/|g_0|^2$ and $\alpha_4=\sigma^2/|g_0|^2$.

The problem of (\ref{MaxSecCapCJoneOP2}) makes sense when the maximum is greater than zero, i.e., when positive secrecy rate is achieved. The conditions under which positive secrecy rate is achieved are given in the following lemma. The proof is given in Appendix \ref{proofLemcondPosRate}.
\begin{lemma}\label{condPosRate}
{\em The condition under which positive secrecy rate is achieved is:
\begin{itemize}
  \item $\alpha_2<\alpha_4$ (i.e., $|h_0|^2>|g_0|^2$);
  \item $\alpha_2>\alpha_4,
P_0>(\alpha_2-\alpha_4)/(\alpha_3G(z_0)-\alpha_1z_0)$ where $z_0$
is the unique root of $\alpha_3G'(z)=\alpha_1$ given by
$z_0=1/(1+u_0^2)$,
$u_0=[\alpha_1/\alpha_3-(2\eta^2-1)+\sqrt{(\alpha_1/\alpha_3)^2+1-2(\alpha_1/\alpha_3)(2\eta^2-1)}\,]/(2\eta\sqrt{1-\eta^2})$.
\end{itemize}
}
\end{lemma}
{\em Remarks}:
The second condition means that if the source-destination channel is weaker than source-eavesdropper channel, direct transmission can not achieve positive secrecy rate, at that time, relays should be used and the total power (source plus relay) should be greater than a threshold.

In the following analysis, we assume the conditions in Lemma \ref{condPosRate} hold. Denote the objective in (\ref{MaxSecCapCJoneOP2}) as $M_1(z)$. It is easy to show that $M_1'(z)>0$. Thus, $z=0$ is not the optimal point. Before proceeding, we give a suboptimal solution which turns out to be the suboptimal solution proposed in \cite{Dong3} which is a special case corresponding to $z=0$ (please see Appendix \ref{proofLemsubopCJmaxSec} for details).
\begin{lemma}
\label{lem:subopCJmaxSec}
{\em When $z=0$ is fixed, a suboptimal solution is given:
\begin{itemize}
  \item if $\alpha_2>\alpha_4$, $P_0<(\alpha_2-\alpha_4)/(\alpha_3(1-\eta^2))$, then
$P_{s,\mathrm{sub}}=0$;
  \item if \{$\alpha_2<\alpha_4$ or
$\alpha_2>\alpha_4$,
$P_0>(\alpha_2-\alpha_4)/(\alpha_3(1-\eta^2))$\} and $(P_0+\alpha_4)\alpha_4>(P_0+\alpha_2)(P_0\alpha_3(1-\eta^2)+\alpha_4)$,
then $P_{s,\mathrm{sub}}=P_0$;
  \item if \{$\alpha_2<\alpha_4$ or
$\alpha_2>\alpha_4$,
$P_0>(\alpha_2-\alpha_4)/(\alpha_3(1-\eta^2))$\} and $(P_0+\alpha_4)\alpha_4 < (P_0+\alpha_2)(P_0\alpha_3(1-\eta^2)+\alpha_4)$,
then $P_{s,\mathrm{sub}}=(-c_3d_2+\sqrt{c_3^2d_2^2-c_3d_2(1-d_2)(\alpha_2-c_3)})/(d_2(1-d_2))$,
where $d_2=\alpha_3(1-\eta^2),
c_3=\alpha_4+P_0\alpha_3(1-\eta^2)$.
\end{itemize}
}
\end{lemma}

Now we proceed. The methodology to solve the problem of (\ref{MaxSecCapCJoneOP2}) is: 1) fix $z$, find the optimal $P_s$; 2) fix $P_s$, find the optimal $z$. Based on this, we  propose an algorithm to search for the optimal $P_s$ and $z$ as follows.
\begin{algorithm}\label{Algo:local}
{\em Take a feasible point $z^{(1)}$ as initial point. Subsequently, find the optimal $P_s^{(1)}$ and then the optimal $z^{(2)}$. Then find the optimal $P_s^{(2)}$, and so on. The procedure converges to the optimal $P_s^\circ$ and $z^\circ$.}
\end{algorithm}

The algorithm \ref{Algo:local} is not complete without providing the methods to find the optimal $z$ for fixed $P_s$ and the optimal $P_s$ for fixed $z$. Next, we provide such methods.
\underline{First}, we consider the problem: find the optimal $P_s$ for fixed $z$. This corresponds to an optimization problem of a single variable $P_s$, and the maximum is achieved at either $0$, $P_0$, or the points with zero derivative. Setting the derivative of the objective to zero leads to the following quadratic equation
\equ{ A_1P_s^2+B_1P_s+C_1=0\label{MaxSecCapquaEq} }
where
$A_1=a_1d_1(1-d_1)-b_1c_1(1-b_1)$, $B_1=2a_1c_1(d_1-b_1)$,
$C_1=a_1c_1(a_1-c_1)$, $a_1=P_0\alpha_1z+\alpha_2$,
$b_1=\alpha_1z$, $c_1=P_0\alpha_3G(z)+\alpha_4$ and
$d_1=\alpha_3G(z)$.
We can obtain $P_s$ explicitly as a function of $z$.
\underline{Second}, we consider the problem: when $P_s$ is fixed, find the optimal $z$. This corresponds to an optimization problem of a single variable $z$, and the maximum is achieved at one of the following points: $0$, $1$ and the points with zero derivative.
The problem to solve can now be rewritten as
\al{
&\max_{z}\ R_s(z)=\log\frac{\alpha_1 z+b_3}{\alpha_1 z+a_3} -\log\frac{\alpha_3G(z)+d_3}{\alpha_3G(z)+c_3}\label{MaxSecCapCJoneOP2fixPs}\\
&\mathrm{s.t.}\quad z\in [0,1]\nnb
}
where
$a_3=\alpha_2/(P_0-P_s)$,
$b_3=(\alpha_2+P_s)/(P_0-P_s)$,
$c_3=\alpha_4/(P_0-P_s)$ and
$d_3=(\alpha_4+P_s)/(P_0-P_s)$.
The derivative of $R_s(z)$ given by
\al{
R_s'(z)\triangleq\frac{\partial R_s}{\partial z}=\frac{P_s}{P_0-P_s}
\left(\frac{\alpha_3G'(z)}{(\alpha_3G(z)+c_3)(\alpha_3G(z)+d_3)}-\frac{\alpha_1}
{(\alpha_1z+a_3)(\alpha_1z+b_3)}\right). \label{equdiffRsz}
}
It is easy to verify that $R_s'(0)>0$, $R_s'(1)<0$. Thus, the optimal $z$ must be the points with zero derivative. Note that $R_s(z)>0$ holds only when $\alpha_1 z+a_3<\alpha_3G(z)+c_3$, which determines an interval $(\underline{z}, \bar{z}) \subset[0,1]$. Here $\underline{z}$ and $\bar{z}$ can be expressed in closed form from the fact: if $G(z)=\beta_1z+\beta_2$ has real roots over $[0,1]$, then its roots can be expressed as $z=1/(1+u_0^2)$ where $u_0=(\eta\sqrt{1-\eta^2}\pm\sqrt{\eta^2(1-\eta^2)+(\beta_2-1+\eta^2)(\eta^2-\beta_1-\beta_2)})/(\beta_2-1+\eta^2)$. With these, we can restrict our attention to the root of (\ref{equdiffRsz}) over $(\underline{z}, \bar{z})$.

To proceed, we need the following result. The proof is given in Appendix \ref{proofprop:Rs}.
\begin{property}\label{prop:Rs} For $R_s(z)$ defined in (\ref{MaxSecCapCJoneOP2fixPs}),
{\em $\frac{\partial^2 R_s}{\partial
z^2}\big\vert_{z'}<0$ for the stationary point $z'\in (\underline{z}, \bar{z})$ (i.e., the point with
$\frac{\partial R_s}{\partial z}\big\vert_{z'}=0$)}.
\end{property}

According to Property \ref{prop:Rs}, we know that the equation (\ref{equdiffRsz}) has a unique root $z'$ such that when $z<z'$, $\frac{\partial R_s}{\partial z}>0$ and when $z>z'$, $\frac{\partial R_s}{\partial z}<0$. This property ensures that the Newton method would be very effective in searching for  $z'$ and would enjoy quadratic convergence.

\subsection{\label{sec:4}Multiple Eavesdroppers}

We now turn to the case of multiple eavesdroppers ($J > 1$).  We restrict our attention to the DF protocol. The  CJ protocol case with multiple eavesdroppers  is a more difficult problem and will be addressed in future work.

Problem \ref{P1} now becomes
\al{
&\max_{P_s,\qw} \ \min_{j\in I} \
\frac{1}{2}\log\frac{\sigma^2+P_s|h_0|^2+\qw^\dagger\qR_h\qw}
{\sigma^2+P_s|g_{0j}|^2+\qw^\dagger\qR_g^j\qw}\label{MaxSecCapDFmop1}\\
&\mathrm{s.t.}\ \ P_s\in [\psmin, P_0],\ \qw^\dagger\qw=P_0-P_s\nnb
}
where $I=\{1,\cdots, J\}$, $\psmin$ is defined as in the case of a single eavesdropper.
By letting $\qw=\sqrt{P_0-P_s}\,\qx$, we can rewrite the above problem as
\al{
&\max_{P_s,\qx} \ \min_{j\in I} \
\frac{1}{2}\log\frac{\sigma^2+P_s|h_0|^2+(P_0-P_s)\qx^\dagger\qR_h\qx}
{\sigma^2+P_s|g_{0j}|^2+(P_0-P_s)\qx^\dagger\qR_g^j\qx}\label{MaxSecCapDFmop1-B}\\
&\mathrm{s.t.}\ \ P_s\in [\psmin, P_0],\ \qx^\dagger\qx=1.\nnb
}

Before proceeding, we give a suboptimal solution which turns out to be the suboptimal solution in \cite{Dong3}. In \cite{Dong3}, if $N\ge J+1$, a suboptimal solution is obtained when an additional constraint is added: $\qw^\dagger\mb{G}=0$ where $\mb{G}=[\mb{g}_1, \mb{g}_2,\cdots,\mb{g}_J]$. This additional constraint means nulling the energy to the eavesdroppers, and this requires $N>J$ so that we have enough degrees of freedom to ensure this is possible. The proof is given in Appendix \ref{proofLemsubopDFmulti}.
\begin{lemma}
\label{lem:subopDFmulti}
{\em When the constraint $\qw^\dagger\mb{G}=0$ is added, a suboptimal solution is obtained as
\equ{
P_{s,\mathrm{sub}}=\left\{\begin{array}{cl}
                            P_0 & \mathrm{if\ } f_2(P_0)>f_2(\psmin); \\
                            \psmin & \mathrm{else}
                          \end{array}
\right.
}
where the function $f_2(P_s)$ is defined by
\equ{
f_2(P_s)=\frac{\sigma^2+P_s|h_0|^2+(P_0-P_s)\|\qE^\dagger\qh\|^2}{\sigma^2+P_s\max_{j\in I}\{|g_{0j}|^2\}}
}
and $\qE$ is the null space of $\mb{G}^\dagger$ with $\qE^\dagger\qE=\qI$.
}
\end{lemma}

Now we proceed. The methodology to solve the problem of (\ref{MaxSecCapDFmop1-B}) is: 1) fix $\qx$, find the optimal $P_s$; 2) fix $P_s$, find the optimal $\qx$. Based on this, we  propose an algorithm to search for the optimal $P_s$ and $\qx$ as follows.
\begin{algorithm}
\label{Algo:DFmulti}
{\em Take a feasible $P_s^{(1)}$ as an initial point. Subsequently, find the optimal $\qx^{(1)}$ and then the optimal $P_s^{(2)}$. Then find the optimal $\qx^{(2)}$, and so on.}
\end{algorithm}
{\em Remarks}: We will see later that: for $J=2$, the procedure always converges to the optimal $P_s$ and $\qx$, while for $J>2$ the procedure does not necessarily converge to the optimal solution. We will discuss this in the sequel.

The algorithm \ref{Algo:DFmulti} is not complete without providing the methods to find the optimal $\qx$ for fixed $P_s$ and find the optimal $P_s$ for fixed $\qx$. Next, we provide such methods.

\begin{figure}[hbtp]
\centering
\includegraphics{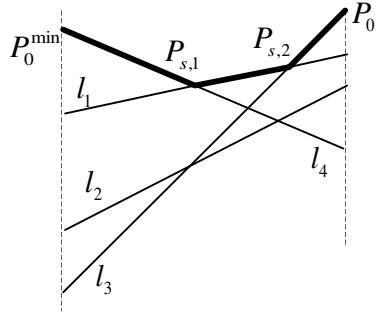}
\caption{Polygonal line for the problem (\ref{MaxSecCapfixX}).} \label{polygonal}
\end{figure}

\underline{First}, we consider the problem: find the optimal $P_s$ for fixed $\qx$. We need to solve
\al{
&\max_{P_s}\
\frac{\sigma^2+P_s|h_0|^2+(P_0-P_s)\qx^\dagger\qR_h\qx}
{\sigma^2+\max_{j\in I}\{P_s|g_{0j}|^2+(P_0-P_s)\qx^\dagger\qR_g^j\qx\}}\label{MaxSecCapfixX}\\
&\mathrm{s.t.}\quad P_s\in [\psmin, P_0].\nnb
}
Note that $\max_{j\in I}\{P_s|g_{0j}|^2+(P_0-P_s)\qx^\dagger\qR_g^j\qx\}$ denotes a polygonal line $\Gamma$ whose vertices are located at the points $P_{s,k}$, $k=0,1,\cdots,M$, $P_{s,0}=\psmin$, $P_{s,M}=P_0$. An example is plotted in Fig. \ref{polygonal}. Note that the objective in (\ref{MaxSecCapfixX}) is a linear fractional function and hence quasi-linear \cite{Boyd} in each line segment $P_{s,k}P_{s,k+1}$, $k=0,1,\cdots,M-1$. Thus, the optimal $P_s$ is one of the vertices of the polygonal, i.e., $P_{s,k}$, $k=0,1,\cdots,M-1$. It is easy to find $P_{s,k}$.

\underline{Second}, we consider the problem: find the optimal $\qx$ for
fixed $P_s$. We need to solve
\al{
&\max_{\qx} \ \min_{j\in I} \
\frac{\sigma^2+P_s|h_0|^2+(P_0-P_s)\qx^\dagger\qR_h\qx}
{\sigma^2+P_s|g_{0j}|^2+(P_0-P_s)\qx^\dagger\qR_g^j\qx}\label{MaxSecCapDFmop1-C}\\
&\mathrm{s.t.}\ \qx^\dagger\qx=1.\nnb
}
We can show that the problem of (\ref{MaxSecCapDFmop1-C}) is equivalent to
\al{
\max_{\qv}\ \|\qv\|^2 \quad\quad \mathrm{s.t.}\
\ \qv^\dagger\qA_j\qv\le 1, j\in I\label{primalDFb}
}
where
\al{
\qA_j&=\wt{\qR}_h^{-1/2}\wt{\qR}_g^j\wt{\qR}_h^{-1/2}, j\in I\\
\wt{\qR}_h&=(\sigma^2+P_s|h_0|^2)\qI+(P_0-P_s)\qR_h, \\
\wt{\qR}_g^j&=(\sigma^2+P_s|g_{0j}|^2)\qI+(P_0-P_s)\qR_g^j, j\in I.
}
Let the solution of (\ref{primalDFb}) be $\qv^\circ$. Then, the solution of (\ref{MaxSecCapDFmop1-C}) is given by $\qx^\circ=\wt{\qR}_h^{-1/2}\qv^\circ/\|\wt{\qR}_h^{-1/2}\qv^\circ\|$. The proof is given in Appendix \ref{proofQCQP}.

The problem of (\ref{primalDFb}) belongs to the quadratic constrained quadratic programs (QCQP) which is in general difficult \cite{Ye}, \cite{Ding}. Denote $\qZ=\qv\qv^\dagger$ which enables us to rewrite it as
\al{
\max_{\qZ}\, \mathrm{Tr}(\qZ) \quad \mathrm{s.t.}\
\ \qZ\succeq 0, \mathrm{rank}(\qZ)=1, \mathrm{Tr}(\qA_j\qZ)\le 1, j\in I.\label{DFsdr0}
}
Dropping the constraint $\mathrm{rank}(\qZ)=1$, the semi-definite relaxation (SDR) of the problem of (\ref{DFsdr0}) is given by
\al{
\max_{\qZ}\, \mathrm{Tr}(\qZ) \quad \mathrm{s.t.}\
\ \qZ\succeq 0, \mathrm{Tr}(\qA_j\qZ)\le 1, j\in I.\label{DFsdr}
}
This is a semi-definite program (SDP) and can be effectively solved by CVX \cite{Grant}.

If the problems  of (\ref{primalDFb}) and (\ref{DFsdr}) achieve the same objective value, we say the SDR of (\ref{DFsdr}) is tight (its solution $\qZ^\circ$ does not necessarily have rank one). Obviously, if the solution $\qZ^\circ$ of (\ref{DFsdr}) has rank one, the SDR (\ref{DFsdr}) is tight and the optimal solution $\qv^\circ$ of the problem of (\ref{primalDFb}) can be obtained by simply eigen-decomposing $\qZ^\circ=\qv^\circ{\qv^\circ}^\dagger$. If $\qZ^\circ$ does not have rank one, the problem for the solution of (\ref{primalDFb}) is in general difficult \cite{Ye}, \cite{Ding}. But for the special case $J=2$, the problem can be solved in polynomial time \cite[\S 2.2]{Ye}, \cite[Lemma 2.2.3]{Ding}. We state it as the following theorem.
\begin{theorem}
\label{th:DFm1Jeq2}
{\em When $J=2$, the SDR (\ref{DFsdr}) is always tight and the optimal solution for the problem of (\ref{primalDFb}) can be constructed in polynomial time.}
\end{theorem}

For $J>2$, if $\qZ^\circ$ does not have rank one, we use Gaussian randomization procedure (GRP) to obtain an approximate solution based on the SDR solution \cite{Luo, Luo2}. In detail, we calculate the eigen-decomposition of $\qZ^\circ=\qU\qD\qU^\dagger$ and generate $\qv_l=\mu_l\qU\qD^{1/2}{\bs \xi}_l$ for $l=1,\cdots, L$, where ${\bs \xi}_l$ is a vector of zero-mean, unit-variance complex circularly symmetric uncorrelated Gaussian random variables, $\mu_l$ is chosen such that $\max_{j\in I} \qv_l^\dagger\qA_j\qv_l=1$ (i.e., the constraint of (\ref{primalDFb}) holds). Suppose
\equ{
l^\star=\arg\, \max_{l} \ \|\qv_l\|^2.
}
Then select $\qv^{(l^\star)}$ as an approximation solution.

\section{Transmit Power Minimization under Secrecy Rate Constraint}\label{PminSecRateCons}

In this section, we address Problems \ref{P3} and \ref{P4}.

\subsection{\label{Pminsec:3sub:1}DF-based protocol}

Problem \ref{P3} is to solve
\al{
&\min_{P_s, \qw}\ P_0=P_s+\|\qw\|^2\label{powerminDFoneOP1}\\
&\mathrm{s.t.}\ \
\frac{\sigma^2+P_s|h_0|^2+\qw^\dagger\qR_h\qw}{\sigma^2+P_s|g_0|^2+\qw^\dagger\qR_g\qw}=4^{R_s^0},\,
P_s\ge \psmin.\nnb
}
We provide a closed form solution to this problem which reveals that the optimal weight vector $\qw^\circ$ is a linear combination of $\qh$ and $\mb{g}$. We give the main result of the problem as a theorem. The proof is given in Appendix \ref{proofTheo1-P0min}.
\begin{theorem}
\label{theorem:1}
{\em When $4^{R_s^0}|g_0|^2-|h_0|^2\ne 0$,
\al{
(P_s^\circ, \qw^\circ)=\left\{\begin{array}{cl}
                               (\zeta, \mb{0})  & \mathrm{if\ } \zeta\ge \psmin, \lambda_2\ge -1 \\
                                (\psmin, \sqrt{\frac{\zeta-\psmin)}{|\lambda_2|}}\,\qu_2) & \mathrm{if\ } \zeta\ge \psmin, \lambda_2< -1 \\
                               (\psmin, \sqrt{\frac{\psmin-\zeta}{\lambda_1}}\,\qu_1)  & \mathrm{if\ } \zeta< \psmin
                              \end{array}
\right.
}
where $\zeta=(4^{R_s^0}-1)\sigma^2/(|h_0|^2-4^{R_s^0}|g_0|^2)$; $\lambda_1>0$, $\lambda_2<0$ are the only two nonzero eigenvalues of $\wt{\qR}=(\qR_h-4^{R_s^0}\qR_g)/(4^{R_s^0}|g_0|^2-|h_0|^2)$ with associated eigenvectors $\qu_1$ and $\qu_2$ (see Lemma \ref{lem:1}).

When $4^{R_s^0}|g_0|^2-|h_0|^2= 0$, the solution is
\al{
P_s^\circ&=\psmin\\
\qw^\circ&=\sqrt{(4^{R_s^0}-1)\sigma^2/\xi_1}\,\qv_1
}
where $\xi_1>0$, $\xi_2<0$ are the only two nonzero eigenvalues of $(\qR_h-4^{R_s^0}\qR_g)$ with associated eigenvectors $\qv_1$ and $\qv_2$ (see Lemma \ref{lem:1}).}
\end{theorem}
\mbox{}

Before ending of this subsection, we should point out that the results in \cite{Dong1}-\cite{Dong3} address only the case $4^{R_s^0}|g_0|^2-|h_0|^2> 0$, thus our analysis here is more complete. Further, our analysis reveals that $(\qR_h-4^{R_s^0}\qR_g)$ has only two nonzero eigenvalues (one positive, and one negative) and, unlike \cite{Dong1}-\cite{Dong3}, provides simple expression for them.

\subsection{\label{Pminsec:3sub:2}CJ-based protocol}

Problem \ref{P4} is to solve
\al{
&\min_{P_s,\qw} \ P_0=P_s+\|\qw\|^2\label{CJoneOP1}\\
&\mathrm{s.t.}\ \
\frac{|h_0|^2}{\qw^\dagger\qR_h\qw+\sigma^2}-\frac{2^{R_s^0}|g_0|^2}{\qw^\dagger\qR_g\qw+\sigma^2}=\frac{2^{R_s^0}-1}{P_s},\,
P_s> 0.\nnb
}
By denoting $\|\qw\|^2=\gamma$, $\qx=\qw/\|\qw\|$, $\qv_1=\qh/\|\qh\|$, $\qv_2=\mb{g}/\|\mb{g}\|$, the problem of (\ref{CJoneOP1}) can be rewritten as
\al{
&\min_{P_s,\qx} \ P_0=P_s+\gamma\label{CJoneOP1-b}\\
&\mathrm{s.t.}\ \
\frac{|h_0|^2}{\gamma\|\qh\|^2\qx^\dagger\qv_1\qv_1^\dagger\qx+\sigma^2}-\frac{2^{R_s^0}|g_0|^2}
{\gamma\|\mb{g}\|^2\qx^\dagger\qv_2\qv_2^\dagger\qx+\sigma^2}=\frac{2^{R_s^0}-1}{P_s},\,
P_s> 0.\nnb
}
Let $\qx$ be a feasible point, and $\qx^\dagger\qv_1\qv_1^\dagger\qx=z$, $z\in [0,1]$. For fixed $z$ and $\gamma$, a larger $\qx^\dagger\qv_2\qv_2^\dagger\qx$ results in a smaller $P_s$. With this and from Lemma \ref{lem:2}, we know that the optimal $\qx^\dagger\qv_2\qv_2^\dagger\qx$ equals $F(z)$ where $F(z)\triangleq 1-(\rho\sqrt{1-z}-\sqrt{(1-\rho^2)z})^2$, $\rho=|\qv_1^\dagger\qv_2|$. With this in mind, we can rewrite (\ref{CJoneOP1}) as
\al{
&\min_{P_s,\, \gamma,\, z} \ P_0=P_s+\gamma\label{CJoneOP3}\\
&\mathrm{s.t.}\quad \frac{1}{\gamma \alpha_1
z+\alpha_2}-\frac{1}{\gamma\alpha_3F(z)+\alpha_4}
=\frac{1}{P_s},\nnb\\
&\quad\quad z\in[0, 1],\,  P_s>0,\, \gamma\ge 0\nnb
}
where
$\alpha_1=\|\qh\|^2(2^{R_s^0}-1)/|h_0|^2$,
$\alpha_2=\sigma^2(2^{R_s^0}-1)/|h_0|^2$,
$\alpha_3=\|\mb{g}\|^2(2^{R_s^0}-1)/(2^{R_s^0}|g_0|^2)$ and
$\alpha_4=\sigma^2(2^{R_s^0}-1)/(2^{R_s^0}|g_0|^2)$.
Further, we can rewrite the problem of  (\ref{CJoneOP3}) as
\al{
&\min_{\gamma,\, z} \ P_0=\frac{(\gamma \alpha_1
z+\alpha_2)[\gamma\alpha_3F(z)+\alpha_4]}{\gamma (\alpha_3F(z)-\alpha_1 z)+\alpha_4 -\alpha_2}+\gamma\label{CJoneOP3-B}\\
&\mathrm{s.t.}\quad \gamma (\alpha_3F(z)-\alpha_1 z)+\alpha_4 -\alpha_2 > 0, \nnb\\
&\quad\quad z\in[0, 1], \gamma\ge 0.\nnb
}

Before proceeding, we give a suboptimal solution which is turns out to be the same as the the suboptimal solution of \cite{Dong3}. Please see Appendix \ref{proofLemsubopsubopCJminP0} for details.
\begin{lemma}\label{lem:subopCJminP0}
{\em When $z=0$ is fixed, a suboptimal solution is pbtained as
\equ{
P_{0,\mathrm{sub}}=P_{s,\mathrm{sub}}+\gamma_{\mathrm{sub}}
}
where
\al{
&\gamma_{\mathrm{sub}}=\left\{\begin{array}{cl}
                                0 & \mathrm{if\ } \frac{\alpha_4}{\alpha_2}>1+\sqrt{\alpha_3(1-\rho^2)} \\
                                \frac{\alpha_2\sqrt{\alpha_3(1-\rho^2)}+\alpha_2-\alpha_4}{\alpha_3(1-\rho^2)}
                                & \mathrm{else}.
                              \end{array}
\right.\\
&P_{s,\mathrm{sub}}=\frac{1}{1/\alpha_2-1/[\gamma_{\mathrm{sub}}\alpha_3(1-\rho^2)+\alpha_4]}.
}
}
\end{lemma}

Now we proceed. The methodology to solve the problem of (\ref{CJoneOP3-B}) is: 1) fix $z$, find the optimal $\gamma$; 2) fix $\gamma$, find the optimal $z$. Based on this, we  propose an algorithm to search for the optimal $\gamma$ and $z$ as follows.
\begin{algorithm}\label{lem:local}
{\em Take a feasible point $z^{(1)}$ as initial point. Subsequently, find the optimal $\gamma^{(1)}$ and then the optimal $z^{(2)}$. Then find the optimal $\gamma^{(2)}$, and so on. The procedure converges to the optimal $\gamma^\circ$ and $z^\circ$.}
\end{algorithm}

The algorithm \ref{lem:local} is not complete without providing the methods to find the optimal $z$ for fixed $\gamma$ and find the optimal $\gamma$ for fixed $z$. Next, we provide such methods.

\underline{First}, we consider the problem: find the optimal $\gamma$ for fixed $z$. This corresponds to an optimization problem of a single variable $\gamma$, and the maximum is achieved at one of the following points: $0$ and the points with zero derivative. With this, we obtain
\begin{equation}
\label{gamma2}
\gamma=\max\left\{\frac{\sqrt{f_1(z)}-(\alpha_4-\alpha_2)}{\alpha_3F(z)-\alpha_1z}, 0\right\}
\end{equation}
where $f_1(z)=(\alpha_2\alpha_3 F(z)-\alpha_1\alpha_4 z)^2/[\alpha_3F(z)-\alpha_1z+\alpha_1\alpha_3zF(z)]$. We can obtain $\gamma$ explicitly as a function of $z$ (in this sense, we in fact reduce the original problem to a single variable optimization $P_0=P_0(z), z\in [0,1]$ explicitly).

\underline{Second}, we consider the problem: when $\gamma$ is fixed, find the optimal $z$. Let us denote the left side of the first constraint in (\ref{CJoneOP3}) by $f_{\gamma}(z)$, namely
\equ{
f_{\gamma}(z)=\frac{1}{\gamma \alpha_1 z+\alpha_2}-\frac{1}{\gamma\alpha_3F(z)+\alpha_4}.\label{f_gamma}
}
We know that the optimal $z$ must maximize $f_{\gamma}(z)$ which results in the minimal $P_s$ (see (\ref{CJoneOP3})). The derivative of $f_{\gamma}(z)$ given by
\equ{
f'_{\gamma}(z)\triangleq\frac{\partial f_{\gamma}}{\partial z}=
\frac{-\gamma \alpha_1}{(\gamma \alpha_1 z+\alpha_2)^2}+\frac{\gamma\alpha_3F'(z)}{(\gamma\alpha_3F(z)+\alpha_4)^2}.
\label{df1root}
}
It is easy to verify that $f_{\gamma}'(0)>0$, $f_{\gamma}'(1)<0$. Thus, $z=0$ is not the optimal point, and
the optimal $z$ must be the points with zero derivative. Note that $P_s>0$ (i.e., $f_{\gamma}(z)>0$) holds only when $\gamma \alpha_1 z+\alpha_2 < \gamma\alpha_3F(z)+\alpha_4$ which determines an interval $(\underline{z}, \bar{z}) \subset[0,1]$. Here $\underline{z}$ and $\bar{z}$ can be expressed in closed form from the fact: if $F(z)=\beta_1z+\beta_2$ has real roots over $[0,1]$, then its roots can be expressed as $z=1/(1+u_0^2)$ where $u_0=(\rho\sqrt{1-\rho^2}\pm\sqrt{\rho^2(1-\rho^2)+(\beta_2-1+\rho^2)(\rho^2-\beta_1-\beta_2)})/(\beta_2-1+\rho^2)$. With these, we can restrict our attention to the root of (\ref{df1root}) over $(\underline{z}, \bar{z})$.

To proceed, we need the following result. The proof is given in Appendix \ref{proofprop:f}.
\begin{property}\label{prop:f}
For $f_{\gamma}(z)$ defined in (\ref{f_gamma}),
{\em $\frac{\partial^2 f_{\gamma}}{\partial
z^2}\big\vert_{z'}<0$ for the stationary point $z'\in (\underline{z}, \bar{z})$ (i.e., the point with
$\frac{\partial f_{\gamma}}{\partial z}\big\vert_{z'}=0$)}.
\end{property}

According to Property \ref{prop:f}, we know that the equation (\ref{df1root}) has a unique root $z'$ such that when $z<z'$, $f_{\gamma}'(z)>0$ and when $z>z'$, $f_{\gamma}'(z)<0$. This property ensures that the Newton method would be very effective in searching for $z'$ and would enjoy quadratic convergence.

\section{\label{sec:5}Numerical Simulations}

In this section we provide some numerical simulations to illustrate the proposed solutions. We use the same system configuration as that in \cite{Dong3}, where source, relays, destination and eavesdroppers are placed along a line. Channels between any two nodes are modeled as a line-of-sight (LOS) channel $\rho_0 d^{-c/2}e^{\mathrm{i}\theta}$, where $d$ is the distance between two nodes, $\rho_0$ is a constant, $c$ is the path loss exponent, and $\theta$ is the phase uniformly distributed within $[0,2\pi)$. In our simulations we set $c=3.5$ and $\rho_0=1$. We assume the distances between relays are much smaller than the distances between relays and source/desination, such that the path loss between different relays and source/destination can be taken as approximately the same.  
Similarly, the path loss between different eavesdroppers and source/destination/relay are approximately the same as well. The results are obtained using Monte-Carlo simulations  consisting of $500$ independent trials.

\begin{figure}[hbtp]
\centering
\includegraphics[width=3in]{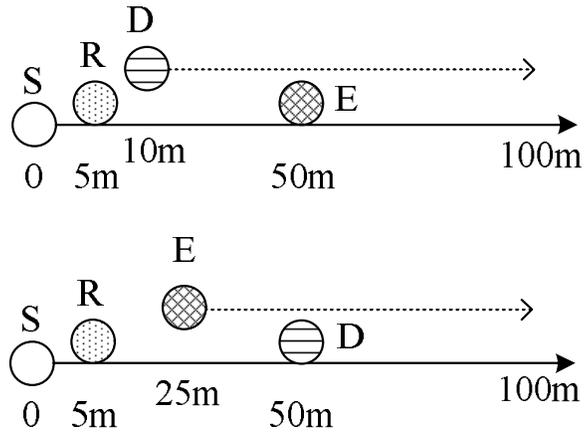}
\caption{Simulation model: source, relays, destination, eavesdroppers
are placed along a line; Upper one for Fig. 4, 5, 7; Lower one for Fig. 8.}\label{fig:sim}
\end{figure}

\begin{figure}[hbtp]
\centering
\includegraphics[width=4in]{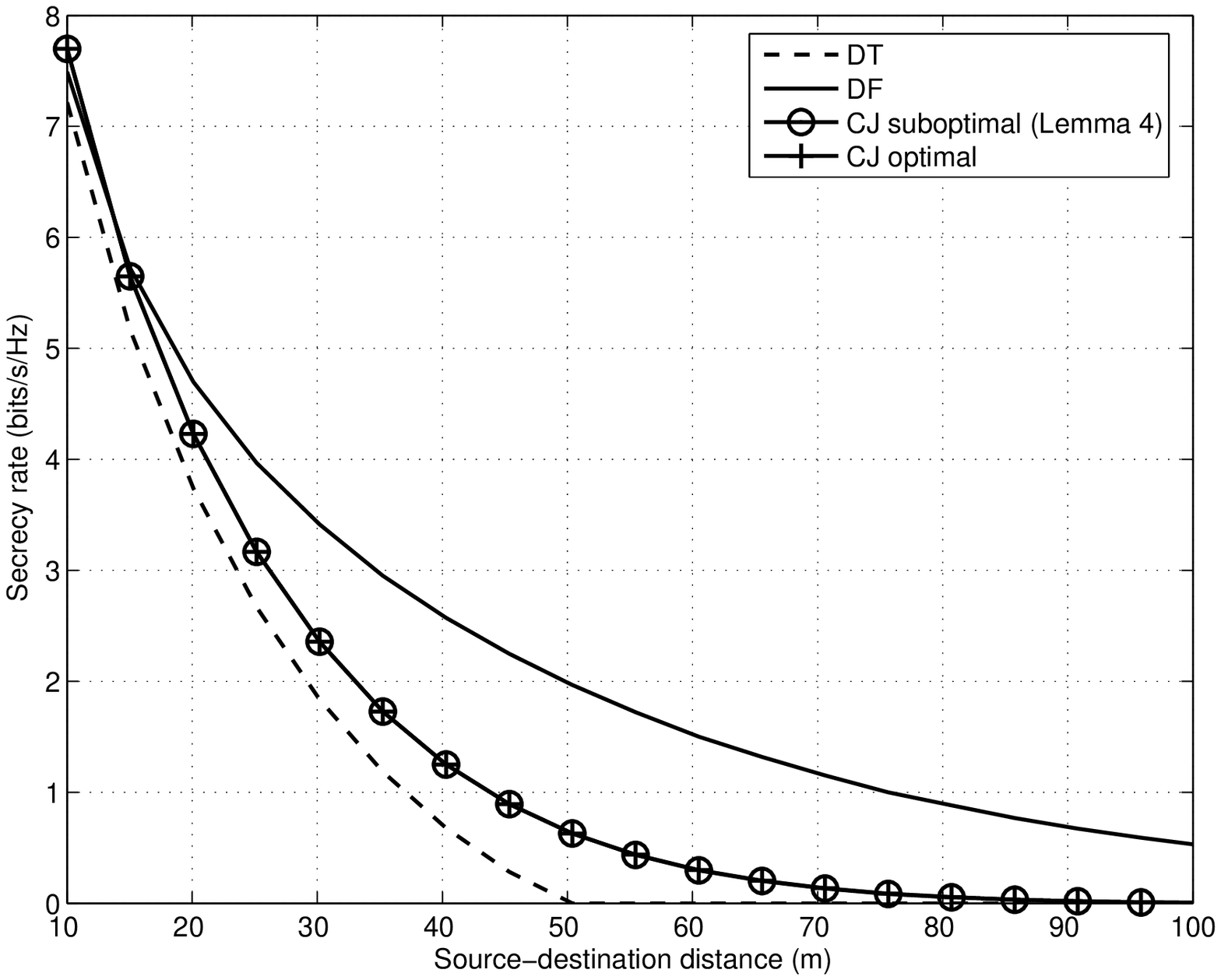}
\caption{Secrecy rate vs. source-destination distance (power constraint: $30\,\mathrm{dBm}$,  source-relay
distance: $5\, \mathrm{m}$,  number of relays:  $N=10$,   one eavesdropper $(J=1)$, source-eavesdropper distance:
 $50\, \mathrm{m}$).} \label{fig:2}
\end{figure}

\begin{figure}[hbtp]
\centering
\includegraphics[width=4in]{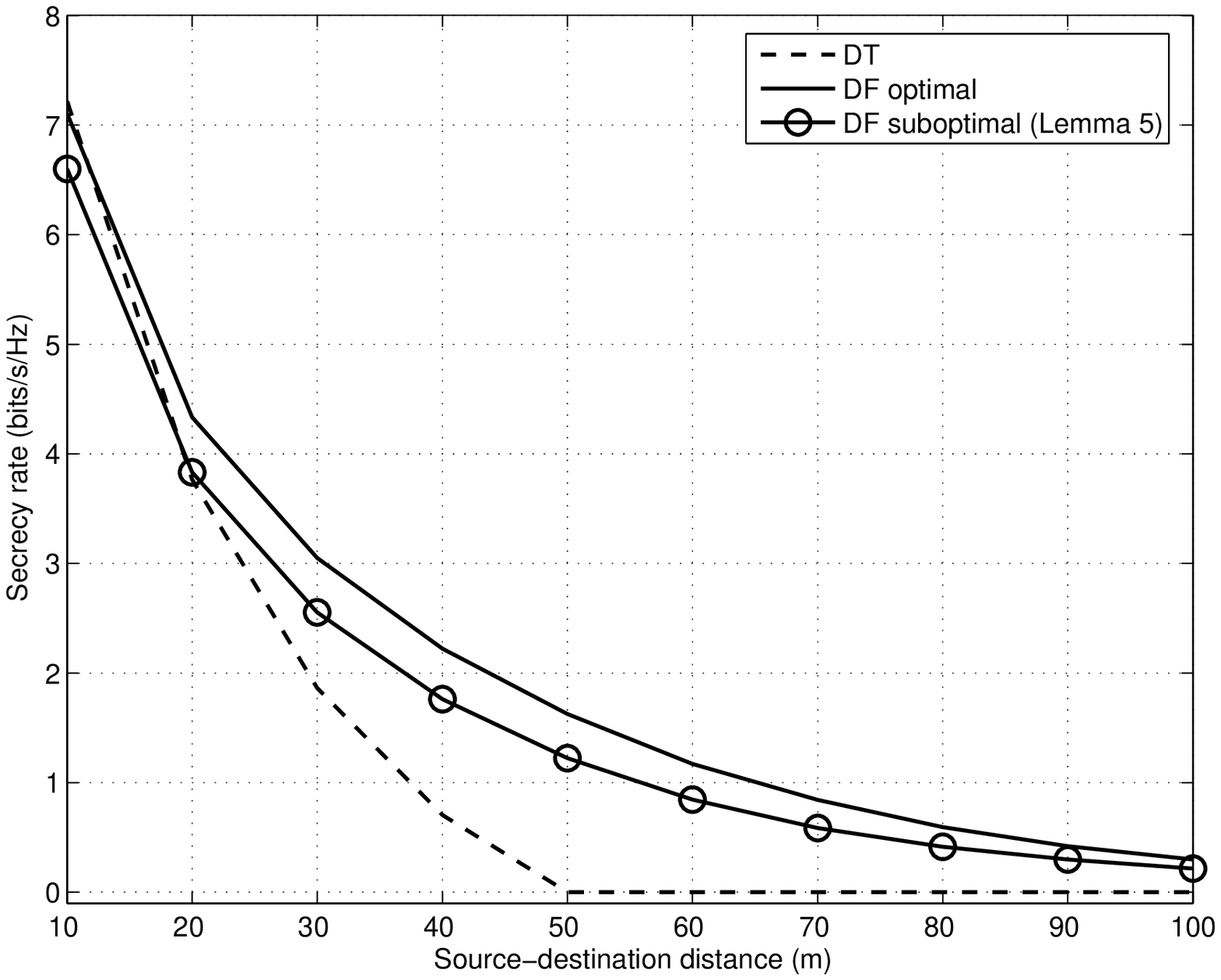}
\caption{Secrecy rate vs. source-destination distance (power constraint: $30\,\mathrm{dBm}$,  source-relay
distance: $5\, \mathrm{m}$,  number of relays:  $N=10$, number of eavesdroppers: $J=7$, source-eavesdropper distance:
 $50\, \mathrm{m}$).} \label{fig:3}
\end{figure}

First, we vary the position of the destination so that the source-destination distance changes from $10\, \mathrm{m}$ to $100\,\mathrm{m}$, as shown in the upper row of Fig. \ref{fig:sim}. The source-relay distances are fixed at $5\,\mathrm{m}$, the number of relays is $N=10$, the source-eavesdropper distances are fixed at $50\, \mathrm{m}$, the power constraint is fixed at $30\,\mathrm{dBm}$, the secrecy rate constraint is fixed at $1\, \mathrm{bits/s/Hz}$. The secrecy rate for a single eavesdropper and multiple eavesdroppers is depicted in Figs. \ref{fig:2} and  \ref{fig:3}, respectively. 
From these two figures, one can see that when the destination moves past the eavesdropper direct transmission cannot sustain positive secrecy rate. On the other hand, both DF and CJ  maintain positive secrecy rate even when the destination is further away from the source than the eavesdropper. The fact that there is a cooperation advantage even when the  destination is at the same location as the eavesdropper is because of the  phases differences of the corresponding channels. Although the propagation environment would be the same for both destination and eavesdropper in that case, the phases will be different due to different receiver phase offsets. The DF scheme yields the higher secrecy rate, while the optimal and suboptimal CJ schemes produce the same average rate.

Similar observations can be drawn from  Fig.  \ref{fig:3} for  the case of multiple eavesdropper as far as the advantage of cooperation over direct transmission is concerned.

\begin{figure}[hbtp]
\centering
\includegraphics[width=4in]{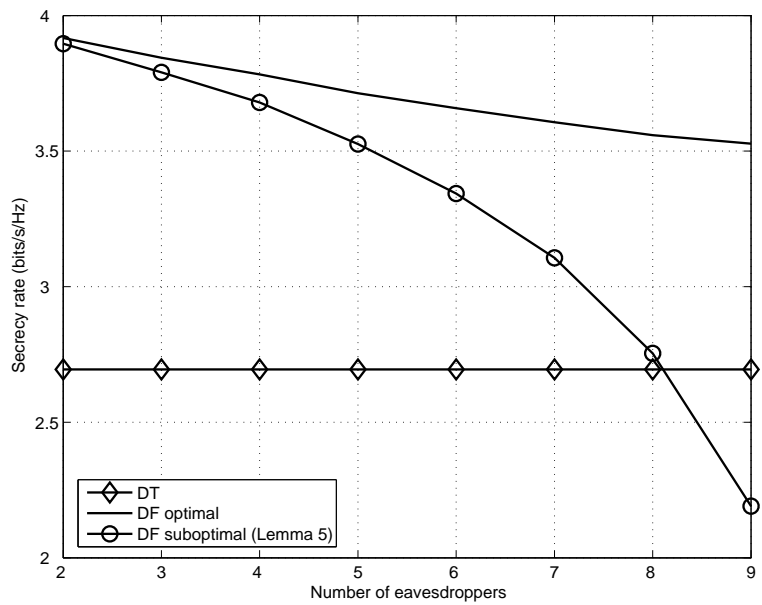}
\caption{Secrecy rate vs. number of eavesdroppers (power constraint: $30\,\mathrm{dBm}$,  source-relay
distance: $5\, \mathrm{m}$, number of relays:  $N=10$, source-destination distance: $25\,\mathrm{m}$, source-eavesdropper distance:
 $50\, \mathrm{m}$).} \label{fig:4}
\end{figure}

\begin{figure}[hbtp]
\centering
\includegraphics[width=4in]{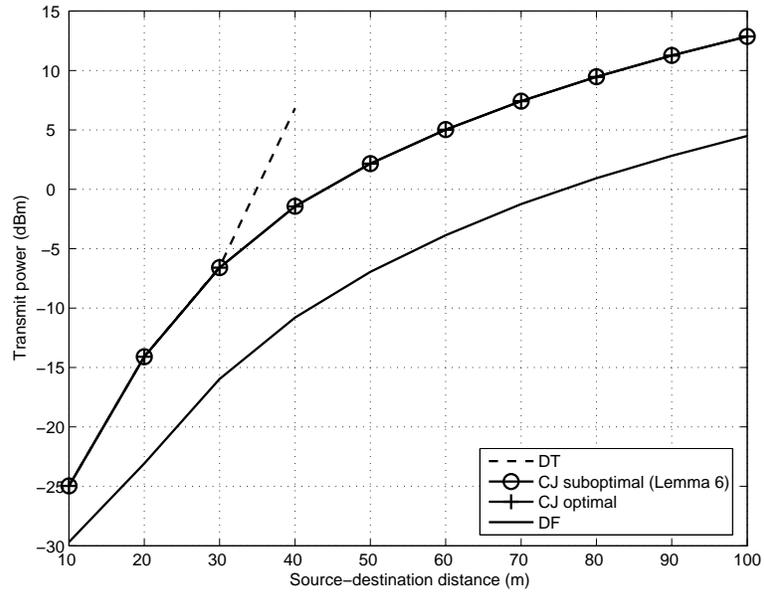}
\caption{Transmit power vs. source-destination distance (secrecy
rate constraint:  $1\, \mathrm{bits/s/Hz}$,  source-relay
distance: $5\, \mathrm{m}$, the number of relays:  $N=10$, one eavesdropper $(J=1)$,  source-eavesdropper distance:
 $50\, \mathrm{m}$).} \label{fig:5}
\end{figure}

In Fig. \ref{fig:4}, the secrecy rate for a fixed configuration and  variable number of eavesdroppers is shown.
It can be seen from Fig. \ref{fig:4} that, when the number of eavesdroppers increases, the suboptimal solution for DF in Lemma \ref{lem:subopDFmulti} becomes inferior as compared to the optimal solution. The minimal transmit power is depicted in Fig. \ref{fig:5}. For comparison purposes, the suboptimal solution for CJ in Lemma \ref{lem:subopCJmaxSec} and the direct transmission result are also shown on the same figure.

\begin{figure}[hbtp]
\centering
\includegraphics[width=4in]{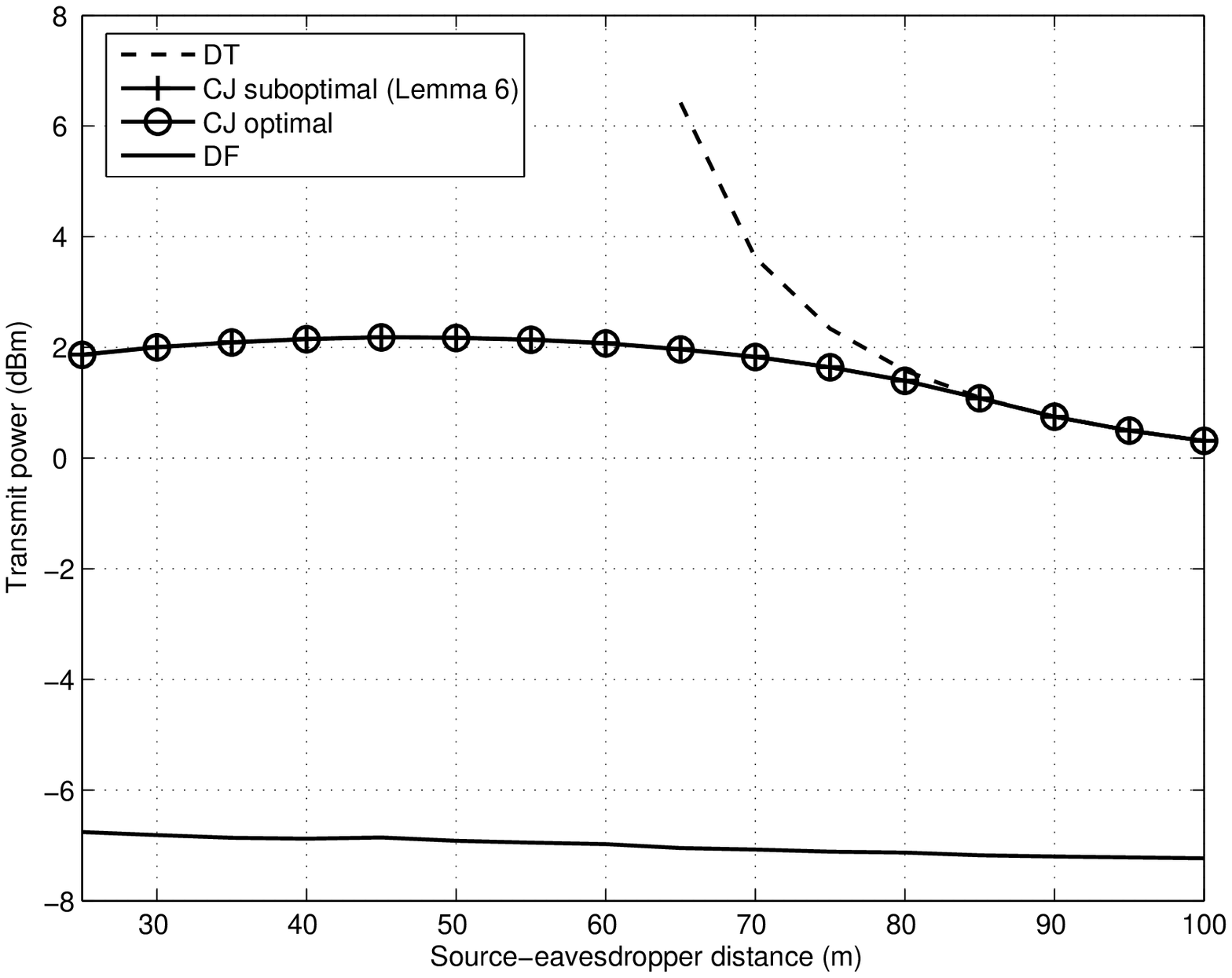}
\caption{Transmit power vs. source-eavesdropper distance (secrecy
rate constraint:  $1\, \mathrm{bits/s/Hz}$, source-relay
distance:  $5\, \mathrm{m}$, number of relays:  $N=10$, one eavesdropper $(J=1)$,  source-destination distance:
 $50\, \mathrm{m}$).} \label{fig:6}
\end{figure}

Second, in Fig. \ref{fig:6}, we show the performance when the eavesdroppers' positions change while the source-destination distance is fixed at $50\,\mathrm{m}$ and the source-relay distances are fixed at $5\,\mathrm{m}$. When the source-eavesdropper distance changes from $25\,\mathrm{m}$ to $100\,\mathrm{m}$ as shown in the lower row of Fig. \ref{fig:sim}, the minimum transmit power for CJ first increases a little, and then decreases, while the minimum transmit power for DF always decreases. The results  show that cooperation can significantly improve the system performance as compared to direct transmission. In particular, when the source-eavesdropper distance is smaller than $65$ m, using direct transmission there is no level of transmit power than can meet the secrecy rate constraint. Also, for source-eavesdropper distance greater than $85$ m direct transmission and the CJ scheme are equivalent in terms of the minimum required transmit power.  The DF approach requires significantly smaller power to meet the secrecy rate constraint. It is interesting to note that in the average sense, the suboptimal solution for CJ in Lemma \ref{lem:subopCJmaxSec} and Lemma \ref{lem:subopCJminP0} is a very good approximation of the optimal solution.

\section{\label{sec:6}Conclusion}

We have given explicit constructions for the optimal relay weights and source transmission power for maximizing the secrecy rate or minimizing the total transmit power (source transmission power plus relay power) under secrecy rate constraint using the DF and CJ protocols in the presence of a single eavesdropper or multiple eavesdroppers. We present numerical results to compare the secrecy rate under our optimal solutions with the secrecy rate under the sub-optimal solutions in \cite{Dong1}-\cite{Dong3}. Numerical results illustrate that cooperation can significantly improve the system performance as compared to direct transmission.

\appendices

\section{Proof of Lemma \ref{lem:1}}\label{proofLem1}

Let $\lambda\ne 0$ be the eigenvalue of $\qr\qr^\dagger-\qs\qs^\dagger$
associated with the eigenvector $\qa$. Thus we have $(\qr\qr^\dagger-\qs\qs^\dagger)\qa=\lambda\qa$
which leads to $\qa=(1/\lambda)[\qr(\qr^\dagger\qa)-\qs(\qs^\dagger\qa)]$.
Thus, $\qa$ has the
form of a linear combination of $\qr$ and $\qs$.
With this, we let $\qa=\pi_1\qr+\pi_2\qs$ where $\pi_1$ and $\pi_2$ will be determined as follows.
Since $\qa$ has unit norm, we have
\equ{
|\pi_1|^2\|\qr\|^2+\pi_1\pi_2^\ast(\qs^\dagger\qr)+\pi_1^\ast\pi_2(\qr^\dagger\qs)+|\pi_2|^2\|\qs\|^2=1.
}
On the other hand, by inserting $\qa=\pi_1\qr+\pi_2\qs$ into $(\qr\qr^\dagger-\qs\qs^\dagger)\qa=\lambda\qa$, we get
\equ{
[\pi_1(\|\qr\|^2-\lambda)+\pi_2(\qr^\dagger\qs)]\qr=[\pi_1(\qs^\dagger\qr)+\pi_2(\|\qs\|^2+\lambda)]\qs.
}
Since $\qr$ and $\qs$ are linearly uncorrelated, we have
\equ{
\pi_1(\|\qr\|^2-\lambda)+\pi_2(\qr^\dagger\qs)=\pi_1(\qs^\dagger\qr)+\pi_2(\|\qs\|^2+\lambda)=0
}
which leads to
\al{
&\lambda^2-(\|\qr\|^2-\|\qs\|^2)\lambda-(\|\qr\|^2\|\qs\|^2-|\qr^\dagger\qs|^2)=0\label{AppLem1eq1}\\
&\pi_2=\frac{\lambda-\|\qr\|^2}{\qr^\dagger\qs}\pi_1.
}
From Cauchy inequality, we get $\|\qr\|^2\|\qs\|^2-|\qr^\dagger\qs|^2>0$. Thus, The equation (\ref{AppLem1eq1})
has a positive root and a negative root. As a result, we obtain $\lambda$ and the corresponding $\pi_1$ and $\pi_2$.

\section{Proof of Lemma \ref{lem:2}}\label{proofLem2}

The solution $\qz^\circ$ of (\ref{lem1op1}) is a linear combination of $\mb{d}_1$
and $\mb{d}_2$, which follows from its optimality condition
$\mb{d}_2\mb{d}_2^\dagger\qz-\mu_1\mb{d}_1\mb{d}_1^\dagger\qz-\mu_2\qz=0$ or further
$\mu_2\qz=(\mb{d}_2^\dagger\qz)\mb{d}_2-(\mu_1\mb{d}_1^\dagger\qz)\mb{d}_1$
where $\mu_1$ and $\mu_2$ are Lagrange multipliers.
Note that
$e^{-\mathrm{i}\theta_2}\qz^\circ=e^{-\mathrm{i}\theta_2}c_1\mb{d}_1+|c_2|\mb{d}_2$ is
also solution of (\ref{lem1op1}), where $\theta_2$ is the argument
of $c_2$. Consequently, we can restrict $c_2\ge 0$.
Inserting $\qz=c_1\mb{d}_1+c_2\mb{d}_2$ into the
constraints and objective, results in
\al{
|c_1|^2+c_2^2+c_1^\ast c_2\mb{d}_1^\dagger\mb{d}_2+c_1c_2\mb{d}_2^\dagger\mb{d}_1&=1\label{c1c2eq1}\\
|c_1|^2+c_2^2|\mb{d}_1^\dagger\mb{d}_2|^2+c_1^\ast c_2\mb{d}_1^\dagger\mb{d}_2+c_1c_2\mb{d}_2^\dagger\mb{d}_1&=q\label{c1c2eq2}\\
\qz^\dagger\mb{d}_2\mb{d}_2^\dagger\qz=1-|c_1|^2(1-|\mb{d}_1^\dagger\mb{d}_2|^2).\label{obj}
} From (\ref{obj}), we need to minimize $|c_1|^2$.
From (\ref{c1c2eq1}) and (\ref{c1c2eq2}), we get $c_2^2(1-|\mb{d}_1^\dagger\mb{d}_2|^2)=1-q$
which leads to $c_2=\sqrt{(1-q)/(1-r^2)}$. By denoting $c_1=|c_1|e^{\mathrm{i}\theta}$ where $\theta$
is the argument of $c_1$, we can rewrite (\ref{c1c2eq1}) as
\equ{
|c_1|^2+|c_1|c_2r (e^{-\mathrm{i}(\phi+\theta)} +e^{\mathrm{i}(\phi+\theta)})+(c_2^2-1)=0.\label{c1c2eq1-b}
}
It is not difficult to show that the optimal $\theta$ given by
\equ{
\theta=\left\{\begin{array}{cc}
                -\phi & \mathrm{if}\ c_2^2-1<0 \\
                \pi-\phi & \mathrm{if}\ c_2^2-1\ge 0
              \end{array}
\right.
}
and the optimal $|c_1|$ is given by
\equ{
|c_1|=\left\{\begin{array}{cc}
                \sqrt{q}-c_2r & \mathrm{if}\ c_2^2-1<0 \\
                c_2r-\sqrt{q} & \mathrm{if}\ c_2^2-1\ge 0
              \end{array}
\right..
}
With these, we obtain the optimal $c_1=(c_2r-\sqrt{q})e^{\mathrm{i}(\pi-\phi)}$.
Further, from (\ref{obj}), we obtain
\equ{
\qz^\dagger\mb{d}_2\mb{d}_2^\dagger\qz=1-(c_2r-\sqrt{q})^2(1-r^2)=1-(r\sqrt{1-q}-\sqrt{(1-r^2)q})^2.
}

\section{Proof of Theorem \ref{solutionP1}}\label{proofTheo-solutionP1}

First, we derive the optimal weight vector $\qw$ for fixed $P_s\ge \psmin$.
By denoting $\qw=\sqrt{P_0-P_s}\,\qx$, we can rewrite the problem of (\ref{MaxSecCapDFoneEaveOP}) as
\al{
&\max_{\qx} \
\frac{1}{2}\log\left(\frac{\sigma^2+P_s|h_0|^2+(P_0-P_s)\|\qh\|^2\qx^\dagger\qu_2\qu_2^\dagger\qx}
{\sigma^2+P_s|g_0|^2+(P_0-P_s)\|\mb{g}\|^2\qx^\dagger\qu_1\qu_1^\dagger\qx}\right)
\label{MaxSecCapDFoneEaveOP-App}\\
&\mathrm{s.t.} \quad \qx^\dagger\qx=1. \nnb
}
Let $\qx$ be a feasible point and $\qx^\dagger\qu_1\qu_1^\dagger\qx=z$, $z\in [0,1]$.
For fixed $z$, a larger $\qx^\dagger\qu_2\qu_2^\dagger\qx$ results in a larger
objective value in the problem of (\ref{MaxSecCapDFoneEaveOP-App}). With this and from Lemma \ref{lem:2}, we know that
$\qx^\dagger\qu_2\qu_2^\dagger\qx$ equals $L(z)$. Thus, the
problem of (\ref{MaxSecCapDFoneEaveOP-App}) can be rewritten as
\al{
&\max_z \ M(z)=
\frac{1}{2}\log\left(\frac{\sigma^2+P_s|h_0|^2+(P_0-P_s)\|\qh\|^2L(z)}{\sigma^2+P_s|g_0|^2+(P_0-P_s)
\|\mb{g}\|^2z}\right)
\label{DFop2}\\
&\mathrm{s.t.} \ \ 0\le z\le 1.\nnb }
This is an optimization problem of a single variable $z$.
It is easy to show that $M'(0)>0$, $M'(1)<0$. Thus, the optimal $z$ must be the points
with zero derivative, i.e., $M'(z)=0$.
As a result, the solution of (\ref{DFop2}), denoted by $z(P_s)$ as a function of $P_s$,
can be expressed in closed form (\ref{zPs}).

Next we consider the optimal $P_s$ and let
it be $P_s^\circ$. We can state that $P_s^\circ$ is also the
solution of the following associated problem
\al{
&\max_{P_s} \
\frac{1}{2}\log\left(\frac{\sigma^2+P_s|h_0|^2+(P_0-P_s)\|\qh\|^2L(z(P_s^\circ))}{\sigma^2+P_s|g_0|^2+(P_0-P_s)
\|\mb{g}\|^2z(P_s^\circ)}\right)
\label{DFoneEaveOP3}\\
&\mathrm{s.t.} \ \ P_s\in [\psmin, P_0].\nnb
}
To see why this is the case, let assume the solution of (\ref{DFoneEaveOP3}) is
$P_s^\prime$ but not $P_s^\circ$.
Denote the objective in (\ref{DFop2}) as $\frac{1}{2}\log(J_1(z,P_s))$.
Note that the objective in (\ref{DFoneEaveOP3}) is
exactly $\frac{1}{2}\log(J_1(z(P_s^\circ),P_s))$. Recall
that for fixed $P_s^\prime$, the solution of
(\ref{DFop2}) is $z(P_s^\prime)$, which leads to
$\frac{1}{2}\log(J_1(z(P_s^\prime),P_s^\prime))\ge \frac{1}{2}\log(J_1(z(P_s^\circ),P_s^\prime))$.
On the other hand, $\frac{1}{2}\log(J_1(z(P_s^\circ),P_s^\circ))\ge
\frac{1}{2}\log(J_1(z(P_s^\prime),P_s^\prime))$ since $P_s^\circ$ is the solution
of (\ref{MaxSecCapDFoneEaveOP}). Combining both gives
$\frac{1}{2}\log(J_1(z(P_s^\circ),P_s^\circ))\ge \frac{1}{2}\log(J_1(z(P_s^\circ),P_s^\prime))$ which
violates the assumption that the solution of (\ref{DFoneEaveOP3}) is
$P_s^\prime$ but not $P_s^\circ$.

Further, $J_1(z,P_s)$ is a
linear fractional function known to be quasi-linear \cite{Boyd},
thus the maximum always occurs at one of the two ends. Thus, the solution
of (\ref{MaxSecCapDFoneEaveOP}) must be $\psmin$ or $P_0$.
With this, we can also obtain the optimal $\qw$ according to Lemma \ref{lem:2}.

\section{Proof of Lemma \ref{condPosRate}}\label{proofLemcondPosRate}

From the objective in the problem of (\ref{MaxSecCapCJoneOP2}),
it should hold that $(P_0-P_s)\alpha_1 z+\alpha_2 < (P_0-P_s)\alpha_3G(z)+\alpha_4$ for some $P_s\in [0,P_0]$, $z \in [0,1]$.
In other words, it should hold $(P_0-P_s)(\alpha_3G(z)-\alpha_1 z)> \alpha_2 - \alpha_4$ for some $P_s\in [0,P_0]$, $z \in [0,1]$.
Denote $K(z)=\alpha_3G(z)-\alpha_1 z$, $z\in [0,1]$. It is easy to verify that:
1) $K(0)=\alpha_3(1-\eta^2)>0$; 2) $K''(z)<0$; 3) $K'(z)\to +\infty$ as $z\to 0$, $K'(1)<0$.
Here $K'(z)$ and $K''(z)$ denote the first- and second- order derivatives, respectively.
With these, first, if $\alpha_2<\alpha_4$, then $(P_0-P_s)K(z) > \alpha_2 - \alpha_4$ for some $P_s\in [0,P_0]$, $z \in [0,1]$ holds
since $K(0)>0$; second, if $\alpha_2>\alpha_4$, we know that $K(z), z\in [0,1]$ achieves its maximum
at $z_0$ with $K'(z_0)=0$ (i.e., the unique root of the equation $\alpha_3G'(z)-\alpha_1=0$),
and $(P_0-P_s)K(z), P_s\in [0,P_0], z\in [0,1]$ achieves its maximum $P_0K(z_0)$, then, the condition should be $P_0K(z_0) > \alpha_2 - \alpha_4$.

\section{Proof of Lemma \ref{lem:subopCJmaxSec}}\label{proofLemsubopCJmaxSec}

When $z=0$ is fixed, the problem of (\ref{MaxSecCapCJoneOP2}) is reduced to
\al{
&\max_{P_s}\ \log\left(1+\frac{P_s}{\alpha_2}\right) -\log\left(1+\frac{P_s}{(P_0-P_s)\alpha_3(1-\eta^2)+\alpha_4}\right)\label{MaxSecCapCJoneOP2App}\\
&\mathrm{s.t.}\quad P_s\in [0,P_0].\nnb
}
This is an optimization problem of a single variable $P_s$,
and the maximum is achieved at one of the following points: $0$, $P_0$ and the points with zero derivative.
After some calculations, the desired results are obtained.

\section{Proof of Property \ref{prop:Rs}}\label{proofprop:Rs}

According to (\ref{MaxSecCapCJoneOP2}), we can write
\equ{
R_s(z)=\log\big(1+\frac{1}{q_1z+q_2}\big)-\log\big(1+\frac{1}{q_3G(z)+q_4}\big)
}
where $q_1=(P_0-P_s)\alpha_1/P_s$, $q_2=\alpha_2/P_s$, $q_3=(P_0-P_s)\alpha_3/P_s$
and $q_4=\alpha_4/P_s$.
It follows from
\al{
\frac{\partial R_s}{\partial z}\big\vert_{z'}&=-\frac{q_1}{(q_1z'+q_2+1)(q_1z'+q_2)}\nnb\\
&\quad+\frac{q_3G'(z')}{(q_3G(z')+q_4+1)(q_3G(z')+q_4)}=0
}
that
\equ{
\frac{(q_3G(z')+q_4+1)(q_3G(z')+q_4)}{(q_1z'+q_2+1)(q_1z'+q_2)}=\frac{q_3G'(z')}{q_1}.\label{appeqdRs}
}
On the other hand, we know
\al{
\frac{\partial^2 R_s}{\partial z^2}\big\vert_{z'}&=\frac{q_1^2(2(q_1z'+q_2)+1)}{[(q_1z'+q_2+1)(q_1z'+q_2)]^2}\nnb\\
&\quad-\frac{(q_3G'(z'))^2(2(q_3G(z')+q_4)+1)}{[(q_3G(z')+q_4+1)(q_3G(z')+q_4)]^2}\nnb\\
&\quad+\frac{q_3G''(z')}{(q_3G(z')+q_4+1)(q_3G(z')+q_4)}.\label{appeqdRs2}
}
Inserting (\ref{appeqdRs}) into (\ref{appeqdRs2}) and using the fact: $G''(z')<0$ and $q_3G(z')+q_4 > q_1z'+q_2$
(since $z'\in (\underline{z}, \bar{z})$)
leads to the desired result.

\section{Proof of Lemma \ref{lem:subopDFmulti}}\label{proofLemsubopDFmulti}

When the constraint $\qw^\dagger\qG=0$ is added, the problem of (\ref{MaxSecCapDFmop1}) is reduced to
\al{
&\max_{P_s,\qw} \
\frac{1}{2}\log\frac{\sigma^2+P_s|h_0|^2+\qw^\dagger\qR_h\qw}
{\sigma^2+P_s\max_{j\in I}\{ |g_{0j}|^2\} }\label{MaxSecCapDFmop1App}\\
&\mathrm{s.t.}\ \ P_s\in [\psmin, P_0],\ \qw^\dagger\qw=P_0-P_s, \qw^\dagger\qG=0.\nnb
}
From $\qw^\dagger\qw=P_0-P_s$ and $\qw^\dagger\qG=0$, we can obtain
$\qw=\sqrt{P_0-P_s}\,\qE\qz$ and $\qz^\dagger\qz=1$ which, when inserted into (\ref{MaxSecCapDFmop1App}), results in
\al{
&\max_{P_s,\qz} \
\frac{1}{2}\log\frac{\sigma^2+P_s|h_0|^2+(P_0-P_s)\qz^\dagger\qE^\dagger\qR_h\qE\qz}
{\sigma^2+P_s\max_{j\in I}\{ |g_{0j}|^2\} }\label{MaxSecCapDFmop1App2}\\
&\mathrm{s.t.}\ \ P_s\in [\psmin, P_0],\ \qz^\dagger\qz=1.\nnb
}
By using the fact $\qz^\dagger\qE^\dagger\qR_h\qE\qz$ achieves its maximum at $\qz=\qE^\dagger\qh/\|\qE^\dagger\qh\|$,
we rewrite the problem of (\ref{MaxSecCapDFmop1App2}) as
\al{
&\max_{P_s} \
\frac{1}{2}\log\frac{\sigma^2+P_s|h_0|^2+(P_0-P_s)\|\qE^\dagger\qh\|^2}
{\sigma^2+P_s\max_{j\in I}\{ |g_{0j}|^2\} }\label{MaxSecCapDFmop1App3}\\
&\mathrm{s.t.}\ \ P_s\in [\psmin, P_0].\nnb
}
The objective in (\ref{MaxSecCapDFmop1App3}) is $\frac{1}{2}\log f_2(P_s)$.
Note that $f_2(P_s)$ is quasi-linear \cite{Boyd},
thus the maximum occurs at one of the two ends, namely, $\psmin$ or $P_0$.

\section{Proof of Equivalent Problem (\ref{primalDFb})}\label{proofQCQP}

From the constraint $\qx^\dagger\qx=1$, we can rewrite $\sigma^2+P_s|h_0|^2=(\sigma^2+P_s|h_0|^2)\qx^\dagger\qx$
and $\sigma^2+P_s|g_{0j}|^2=(\sigma^2+P_s|g_{0j}|^2)\qx^\dagger\qx$ which enables us to rewrite
the problem of (\ref{MaxSecCapDFmop1-C}) as
\al{
\max_{\|\qx\|=1}\,\min_{j\in I}\,
\frac{\qx^\dagger\wt{\qR}_h\qx}{\qx^\dagger\wt{\qR}_g^j\qx}.\label{MaxSecCapop3}
}
Further, by denoting $\qu=\wt{\qR}_h^{1/2}\qx/\|\wt{\qR}_h^{1/2}\qx\|$, we can rewrite the problem of (\ref{MaxSecCapop3}) as
\al{
\max_{\qu}\min_{j\in I}\,
\frac{1}{\qu^\dagger\qA_j\qu}\quad\quad\mathrm{s.t.}\ \
\qu^\dagger\qu=1.\label{slackP}
}
By introducing the slack variable $y$ to rewrite $\min_{j\in I}
1/(\qu^\dagger\qA_j\qu)=1/y$, we can rewrite the problem of
(\ref{slackP}) as
\al{
\max_{\qu,\ y}\ \frac{1}{y} \quad\quad \mathrm{s.t.}\
\ \qu^\dagger\qu=1,\, \qu^\dagger\qA_j\qu\le y, j\in I.\label{primalDF}
}
Let $\qv=\qu/\sqrt{y}$, then $\qu^\dagger\qu=1$ is equivalent to $\|\qv\|^2=1/y$
and we can rewrite the problem of (\ref{primalDF}) as the problem of (\ref{primalDFb}).
Let the solution of (\ref{primalDFb}) be $\qv^\circ$, then the solution of (\ref{slackP})
is $\qu^\circ=\qv^\circ/\|\qv^\circ\|$, the solution of (\ref{MaxSecCapop3}) is $\qx^\circ=\wt{\qR}_h^{-1/2}\qv^\circ/\|\wt{\qR}_h^{-1/2}\qv^\circ\|$.

\section{Proof of Theorem \ref{theorem:1}}\label{proofTheo1-P0min}

The first constraint of (\ref{powerminDFoneOP1}) leads to \al{
(4^{R_s^0}|g_0|^2-|h_0|^2)P_s=\qw^\dagger(\qR_h-4^{R_s^0}\qR_g)\qw-(4^{R_s^0}-1)\sigma^2.\label{firstCon}
} There are two cases. If $4^{R_s^0}|g_0|^2-|h_0|^2\ne 0$, it
follows from (\ref{firstCon}) that
$P_s=\qw^\dagger\wt{\qR}\qw+\zeta$. With this, we can rewrite the
problem of (\ref{powerminDFoneOP1}) as \al{ \min_{\qw} \
\qw^\dagger\qw+\qw^\dagger\wt{\qR}\qw+\zeta\quad\mathrm{s.t.}\
\qw^\dagger\wt{\qR}\qw+\zeta\ge \psmin.\label{powerminDFoneOP2} }
The optimal relay weight vector $\qw^\circ$ has the form of
$(d_1\qu_1+d_2\qu_2)$ which follows from its optimality condition
of (\ref{powerminDFoneOP2}) \cite{Boyd} \equ{
\qw+\wt{\qR}\qw-\nu\wt{\qR}\qw=0.\label{optConDFone} } where
$\nu\ge 0$ is the Lagrange multiplier. Indeed, inserting
$\wt{\qR}=\lambda_1\qu_1\qu_1^\dagger+\lambda_2\qu_2\qu_2^\dagger$
into (\ref{optConDFone}) gives
$\qw=(\nu-1)\lambda_1(\qu_1^\dagger\qw)\qu_1+(\nu-1)\lambda_2(\qu_2^\dagger\qw)\qu_2$.
With this, we can rewrite (\ref{powerminDFoneOP2}) as \al{
&\min_{d_1, d_2} \ (1+\lambda_1)|d_1|^2+(1+\lambda_2)|d_2|^2+\zeta\label{powerminDFoneOP3}\\
&\mathrm{s.t.}\ \ \lambda_1|d_1|^2+\lambda_2|d_2|^2+\zeta\ge
\psmin.\nnb } There are several case. If $\zeta\ge \psmin$,
$\lambda_2\ge -1$, then the solution is $|d_1|=0$, $|d_2|=0$; If
$\zeta\ge \psmin$, $\lambda_2< -1$, then the solution is
$|d_1|=0$, $|d_2|^2=(\zeta-\psmin)/|\lambda_2|$; If $\zeta<
\psmin$, the solution is $|d_1|^2=(\psmin-\zeta)/\lambda_1$,
$|d_2|=0$.

Similarly, we can obtain the other results in Theorem
\ref{theorem:1}.

\section{Proof of Lemma \ref{lem:subopCJminP0}}\label{proofLemsubopsubopCJminP0}

When $z=0$ is fixed, the problem of (\ref{CJoneOP3-B}) is reduced to
\al{
&\min_{\gamma} \ P_0=\frac{\alpha_2[\gamma\alpha_3(1-\rho^2)+\alpha_4]}{\gamma\alpha_3(1-\rho^2)+\alpha_4-\alpha_2}+\gamma\label{CJoneOP3-BApp}\\
&\mathrm{s.t.}\quad \gamma > \frac{\alpha_2-\alpha_4}{\alpha_3(1-\rho^2)},\gamma\ge 0.\nnb
}
This is an optimization problem of a single variable $\gamma$, and
the maximal is achieved at one of the following points: $0$ and
the points with zero derivative.
After some calculations, the desired results are obtained.

\section{Proof of Property \ref{prop:f}}\label{proofprop:f}

It follows from
\equ{
\frac{\partial f_{\gamma_2}}{\partial z}\big\vert_{z'}=\frac{-\gamma \alpha_1}{(\gamma \alpha_1 z'+\alpha_2)^2}
+\frac{\gamma\alpha_3F'(z')}{(\gamma\alpha_3F(z')+\alpha_4)^2}=0
}
that
\equ{
\frac{(\gamma\alpha_3F(z')+\alpha_4)^2}{(\gamma \alpha_1 z'+\alpha_2)^2}=\frac{\alpha_3F'(z')}{\alpha_1}.\label{eqdf}
}
On the other hand, we know
\al{
\frac{\partial^2 f_{\gamma_2}}{\partial z^2}\big\vert_{z'}&=\frac{2(\gamma \alpha_1)^2}{(\gamma \alpha_1 z'+\alpha_2)^3}
-\frac{2(\gamma\alpha_3F'(z'))^2}{(\gamma\alpha_3F(z')+\alpha_4)^3}\nnb\\
&\quad+\frac{\gamma\alpha_3F''(z')}{(\gamma\alpha_3F(z')+\alpha_4)^2}.\label{eqdf2}
}
Inserting (\ref{eqdf}) into (\ref{eqdf2}) and using the fact: $F''(z')<0$ and
$\gamma\alpha_3F(z')+\alpha_4>\gamma \alpha_1 z'+\alpha_2$ (since $z'\in (\underline{z}, \bar{z})$) leads to the desired result.


\begin{thebibliography}{1}

\bibitem{LiConf1} Jiangyuan~Li, A.~P.~Petropulu, and S.~Weber, ``Transmit power minimization under secrecy
capacity constraint in cooperative wireless
communications, ''in {\em Proc. 15th IEEE Workshop on Statistical Signal Processing}, Cardiff, Wales, UK, Sept. 2009.

\bibitem{LiConf2} Jiangyuan~Li, A.~P.~Petropulu, and S.~Weber, ``Secrecy rate optimization under cooperation with perfect channel state information, ''in {\em Proc. 43rd Annual Asilomar Conference on Signals, Systems, and Computers}, Pacific Grove, CA, Nov. 2009.

\bibitem{Liang} Y.~Liang, H.~V.~Poor, and S.~Shamai (Shitz), ``Secure communication over fading channels,''
{\em IEEE Trans. Inf. Theory}, vol. 54, no. 6, pp. 2470-2492, Jun. 2008.

\bibitem{Gopala} P.~K.~Gopala, L.~Lai, and H.~El~Gamal, ``
On the secrecy capacity of fading channels,''
{\em IEEE Trans. Information Theory}, vol. 54, no. 10, pp. 4687-4698, Oct. 2008.

\bibitem{Liang4} Y.~Liang, H.~V.~Poor, and S.~Shamai (Shitz), ``Physical layer security in broadcast
networks,'' {\em Security and Communication Networks}, vol. 2, pp. 227-238, Wiley, May-Jun. 2009.

\bibitem{Liu} R.~Liu, T.~Liu, H.~V.~Poor, and S.~Shamai (Shitz), ``MIMO Gaussian broadcast
channels with confidential messages, '' in {\em Proceedings
of the IEEE International Symposium on Information Theory (ISIT)}, Seoul,
Korea, June-July 2009.

\bibitem{Liang:book} Y.~Liang, H.~V.~Poor, and S.~Shamai (Shitz),
{\em Information Theoretic Security}, Now Publishers, Delft, The Netherlands, 2009.

\bibitem{Wyner} A.~D.~Wyner, ``The wire-tap channel, '' {\em Bell System Technical Journal}, vol. 54, no. 8, pp. 1355-1387, 1975.

\bibitem{Hellman} S.~K.~Leung-Yan-Cheong and M.~E.~Hellman, ``The Gaussian wire-tap
channel,'' {\em IEEE Trans. Information Theory}, vol. 24, pp. 451-456, Jul. 1978.

\bibitem{Khisti} A.~Khisti and G.~Wornell, ``The MIMOME channel, ''
in {\em Proceedings of the 45th Annual Allerton Conference on Communication,
Control and Computing}, Monticello, IL, USA, Sept. 2007.

\bibitem{Khisti2}
A.~Khisti and G.~W.~Wornell, ``Secure transmission with multiple
antennas: The MISOME wiretap channel,'' {\em IEEE Trans. Inf.
Theory}, Submitted in Aug. 2007. [Online]. Available:
http://arxiv.org/abs/0708.4219

\bibitem{TieLiu}
T.~Liu and S.~Shamai, ``A note on the secrecy capacity of the
multi-antenna wiretap channel,'' {\em IEEE Trans. Inf. Theory},
Submitted in Nov. 2007. [Online]. Available:
http://arxiv.org/abs/0710.4105

\bibitem{Hassibi} F.~Oggier and B.~Hassibi, ``The secrecy capacity of the MIMO wiretap
channel, '' in {\em IEEE International
Symposium on Information Theory (ISIT)}, pp. 524-528, Toronto, ON, Canada, Jul. 2008.

\bibitem{Oggier}
F.~Oggier and B.~Hassibi, ``The secrecy capacity of the MIMO wiretap
channel,'' {\em IEEE Trans. Inf. Theory}, Submitted in Oct. 2007.
[Online]. Available: http://aps.arxiv.org/abs/0710.1920, updated at Jul. 2009.

\bibitem{Bhargava} Z.~Rezki, F.~Gagnon, and V.~Bhargava, ``The ergodic capacity of the MIMO wire-tap channel, ''
[online]. Available: http://arxiv.org/abs/0902.0189v1, Feb. 2009.

\bibitem{Tekin} E.~Tekin and A.~Yener, ``The general Gaussian multiple access and two-way wire-tap channels: achievable rates and cooperative jamming,'' {\em IEEE Trans. Inf. Theory}, vol. 54, no. 6, pp. 2735-2751, Jun. 2008.

\bibitem{Lai} L.~Lai and H.~El~Gamal, ``The relay-eavesdropper channel: cooperation for secrecy,'' {\em IEEE Trans. Inf. Theory}, vol. 54, no. 9, pp. 4005-4019, Sept. 2008.

\bibitem{Tang} X.~Tang, R.~Liu, P.~Spasojevic, and H.~V.~Poor, ``The Gaussian wiretap channel with a helping interferer,'' in {\em Proc. ISIT}, Toronto, Ontario, Canada, Jul. 2008.

\bibitem{Yuksel} M.~Yuksel and E.~Erkip, ``The relay channel with a wire-tapper, ''in {\em Proc.
41st Annual Conference on Information Sciences
and Systems}, Baltimore, MD, Mar. 2007.

\bibitem{Yuksel:ITW07}
M.~Yuksel and E.~Erkip, ``Secure communication with a relay helping
the wiretapper,'' in {\em Proc. 2007 IEEE Information Theory
Workshop}, Lake Tahoe, CA, Sept. 2007.

\bibitem{Aggarwal} V.~Aggarwal, L.~Sankar, A.~R.~Calderbank, and H.~V.~Poor, ``Secrecy capacity
of a class of orthogonal relay eavesdropper channels,'' {\em EURASIP Journal on
Wireless Communications and Networking, Special Issue on Wireless Physical
Layer Security}, to appear.

\bibitem{Dong1} L.~Dong, Z.~Han, A.~Petropulu, and H.~V.~Poor, ``Secure wireless communications via cooperation,'' in {\em 46th Annual Allerton Conference on Communication, Control, and Computing}, pp. 1132-1138, Sept. 2008.

\bibitem{Dong2} L.~Dong, Z.~Han, A.~Petropulu, and H.~V.~Poor, ``Amplify-and-forward based cooperation for secure wireless communications,'' in {\em Proc. ICASSP}, Taipei, Taiwan, Apr. 2009.

\bibitem{Dong3} L.~Dong, Z.~Han, A.~Petropulu, and H.~V.~Poor, ``Improving wireless physical layer security
via cooperating relays,'' {\em IEEE Trans. Signal Processing}, accepted in 2009.

\bibitem{Bloch} M.~Bloch, J.~O.~Barros, M.~R.~D.~Rodrigues, and S.~W.~McLaughlin, ``Wireless information-theoretic security,''
{\em IEEE Trans. Inf. Theory}, vol. 54, no. 6, pp. 2515-2534, Jun. 2008.

\bibitem{Liang:Allerton07} Y.~Liang, G.~Kramer, H.~V.~Poor, and S.~Shamai (Shitz), ``Compound wire-tap channels,'' in {\em Proc. 45th Annual Allerton Conference on Communication, Control, and Computing}, Monticello, IL, USA, Sept. 2007.

\bibitem{Boyd} S.~Boyd and L.~Vandenberghe, ``Convex Optimization,'' {\em Cambridge University Press}, Cambridge, UK, 2004.

\bibitem{Luo} V.~H.~Nassab, S.~Shahbazpanahi, A.~Grami, and Z.~Q.~Luo, ``Distributed beamforming for relay networks
based on second-order statistics of the
channel state information,''
{\em IEEE Trans. Signal Processing}, vol. 56, no. 9, pp. 4306-4316, Sep. 2008.

\bibitem{Luo2} N.~D.~Sidiropoulos, T.~N.~Davidson, and Z.~Q.~Luo, ``Transmit beamforming for physical-layer multicasting,''
{\em IEEE Trans. Signal Processing}, vol. 54, no. 6, pp. 2239-2251, Jun. 2006.


\bibitem{Ye} Y.~Ye and S.~Zhang, ``New results on quadratic minimization, ''
{\em SIAM Journal on Optimization}, vol. 14, no. 1, pp. 245-267, 2003.

\bibitem{Ding} Y.~Ding, ``On efficient semidefinite relaxations for quadratically constrained quadratic programming,''
Master {\em thesis}, University of Waterloo, 2007.

\bibitem{Grant} M.~Grant, S.~Boyd, cvx Users' Guide, 2009.







\end{thebibliography}
\end{document}